\documentclass[amsmath,amssymb,aps,prd,11pt,tightenlines, 
nofootinbib,preprintnumbers,notitlepage]{revtex4-2}
\usepackage{graphicx}
\usepackage{dcolumn}
\usepackage{bm}
\usepackage{amssymb}
\usepackage{amsmath}
\usepackage{mathrsfs}
\usepackage{graphicx,color} 
\usepackage[english]{babel}     
\usepackage{graphicx,wrapfig,lipsum}      
\usepackage{multimedia}    
\usepackage[mathscr]{eucal}  
\usepackage{bm}              
\usepackage{amssymb}            
\usepackage{amsmath}     
\usepackage{mathrsfs}  
\usepackage[mathscr]{eucal}      
\usepackage[normalem]{ulem} 
\usepackage{float}
\usepackage{subfig}
\usepackage{booktabs}
\begin{document} 
\title{From center-vortex ensembles to the confining flux tube}
\author{David R. Junior$^{1,2}$}
\author{Luis E. Oxman$^1$}
\author{Gustavo M. Sim\~oes$^1$}
\affiliation{$^1$Instituto de Física, Universidade Federal Fluminense,\\ Avenida Litorânea s/n, 24210-340 Niterói, Brazil.\\
$^2$Institut f\"ur Theoretische Physik, Universit\"at T\"ubingen,\\ Auf der Morgenstelle 14, 72076 Tübingen, Germany.}
\date{\today} 

\begin{abstract}
In this review, we discuss the present status of the description of confining flux tubes in SU(N) pure Yang-Mills theory in terms of ensembles of percolating center vortices. This is based on three main pillars: modelling in the continuum the ensemble components detected in the lattice, the derivation of effective field representations, and contrasting the associated properties with Monte Carlo lattice results. The integration of the present knowledge about these points is essential to get closer to a unified physical picture for confinement. Here, we shall emphasize the last advances, which point to the importance of including the nonoriented center-vortex component and non-Abelian degrees when modelling the center-vortex ensemble measure. These inputs are responsible for the emergence of topological solitons and the possibility of accommodating the asymptotic scaling properties of the confining string tension.
\end{abstract}
\maketitle
\section{Introduction}

\label{intro}

Our knowledge about the elementary particles,  as well as three of the four known fundamental interactions, is successfully described by the standard model of particle physics. 
In particular, the quantitative behavior of the electromagnetic, weak, and strong interactions is encoded in the common language of gauge theories.  In the strong sector, an important and intriguing phenomenon regarding the possible asymptotic particle states takes place. 
When quarks and gluons are created in a  collision,  they cannot move apart. Instead, they give rise to jets of colorless particles (hadrons) formed by confined quark and gluon degrees of freedom.  
 Although confinement is key for the existence of protons and neutrons, a first-principles understanding of the mechanism underlying this phenomenon is still lacking. At  high energies,  the detailed scattering  properties between quarks and gluons are successfully reproduced by QCD perturbative calculations in the continuum, which are possible thanks to asymptotic freedom. 
 This is in contrast with the status at low-energies, where the validity of quantum chromodynamics (QCD) is well-established from computer simulations of the hadron spectrum which successfully make contact with the observed masses. This review focuses on this type of non-perturbative problem in pure $SU(N)$ Yang--Mills (YM) theory, which is a challenging open problem in contemporary physics. Here again, Monte Carlo simulations provide a direct way to deal with the large quantum fluctuations 
and compute averages of observables such as the Wilson loop,  which is an order parameter for confinement in pure YM theories. As usual, the lattice calculations, as well as the  center-vortex ensembles we shall discuss, consider an Euclidean (3d or 4d) spacetime. Unless explicitly stated, this is the metric that will be used throughout this work.
For heavy quark probes in an irreducible  representation ${\rm D}$, the Wilson loop is given by:
\begin{gather}  
{\mathcal W}_{\rm D} ( {\cal C}_{\rm e})    =  \frac{1}{\mathscr{D}} \, {\rm tr}\, {\rm D}  \left( P \left\{ e^{i \int_{{\cal C}_{\rm e}} dx_\mu\,  A_{\mu}(x)  } \right\} \right) \;,
\label{wloope}
\end{gather} 
where $\mathscr{D}$ is the dimensionality of ${\rm D}$.  
The closed path $\mathcal{C}_{\rm e}$  can be thought of as associated to the creation, propagation, and annihilation of a pair of quark/antiquark probes. From a rectangular path with sides $T$ and $R$,   information about the static interquark potential was obtained from the large $T$ behavior $ \langle {\mathcal W}_{\rm D}(\mathcal{C}_{\rm e})  \rangle \sim e^{-T\,  V_{\rm D}(R)}$.  An area law, given by the propagation time $T$ multiplied by the interquark distance $R$,  corresponds to a linear confining potential~\cite{Wilson} (for a review, see ~\cite{Bali}).

 There are many model-independent facts that point to the importance of the center of the group $SU(N)$ to describe the confining properties of YM theory. In this regard, the first ideas relating the possible phases to the  $Z(N)$ properties of the vacuum
were developed in~\cite{thooft}. There, disorder vortex field and string field operators were introduced in (2~+~1)d and (3~+~1)d Minkowski spacetime, respectively. At equal time,
they satisfy
\begin{gather}
\hat{{\mathcal W}}_{\rm F} ( {\cal C}_{\rm e}) \,  \hat{V}({\bf x }) = e^{i 2\pi \,  L (\mathbf{x}, {\cal C}_{\rm e})/N}\, \hat{V}({\bf x}) \, \hat{{\mathcal W}}_{\rm F} ( {\cal C}_{\rm e}) \;, 
\makebox[.9in]{{\rm in (2~+~1)d,}}
\label{Vop3}
\end{gather}    
\begin{gather}
\hat{{\mathcal W}}_{\rm F} ( {\cal C}_{\rm e}) \, \hat{V}({\cal C}) = e^{i 2\pi\, L(\mathcal{C}, {\cal C}_{\rm e})/N}\, \hat{V}({\cal C}) \, \hat{{\mathcal W}}_{\rm F} ( {\cal C}_{\rm e})  \;, 
\makebox[.9in]{{\rm in (3~+~1)d,}} 
\end{gather} 
where the subindex ${\rm F}$ denotes the fundamental representation, $\mathbf{x} \in \mathbb{R}^2$  ($\mathcal{C} \in \mathbb{R}^3$) is a point (curve) in real space where a thin pointlike (looplike) thin center vortex is created in three (four) dimensional spacetime. $L(\mathbf{x}, {\cal C}_{\rm e} )$ and $L({\cal C},  {\mathcal C}_{\rm e} )$  are 
 the corresponding linking numbers. An explicit realization of $\hat{V}$ was given by the action $ \hat{V} |A\rangle = |A^{S} \rangle $, where  $| A\rangle$ are quantum states with well-defined shape $A_0=0, A_i$ ($i=1,2,3$) at a given time.  The field $A_\mu^S$ has the form of a gauge transformation, but performed with a singular phase $S\in SU(N)$.  To define the operator $\hat{V}(\mathbf{x})$ (respectively $\hat{V}(\mathcal{C})$),  $S$ must change by a center element when going around any spatial closed loop that links ${\bf x}$ (respectively $\mathcal{C}$). Spurious singularities may be eliminated by using the adjoint representation ${\rm Ad}(S)$, which leaves a physical effect only at the point ${\bf x}$, or closed path  $\mathcal{C}$,  where  ${\rm Ad}(S)$ is multivalued. Arguments in favor of characterizing confinement as a {\it magnetic}  $Z(N)$ spontaneous  symmetry breaking  phase \mbox{(center-vortex condensate),}   
\begin{gather}
\langle \hat{V}({\bf x})\rangle   \neq 0  \makebox[.5in]{,}  
\langle \hat{V}({\cal C})\rangle \sim e^{-\mu  {\rm Perimeter} ({\cal C})}   \;, 
\label{criteria}
\end{gather}   
were also  given in that work.   

 The lattice also provides direct information about the role played by the center of $SU(N)$ in the confinement/deconfinement phase transition. This is observed in the properties of the Polyakov loops $P_{\mathbf{x}}(\mathscr{A})$, which  are given by Equation \eqref{wloope} computed on a straight path located at a spatial coordinate $\mathbf{x} $ and extending along the Euclidean time-direction. Due to the finite-temperature periodicity conditions, these segments can be thought of as circles. By considering the fundamental representation,  $  P_{\mathbf{x}}(\mathscr{A})$ was analyzed in the lattice~\cite{STOKES2014341}. 
When changing from higher to lower temperatures, the distribution of the phase factors of $  P_{\mathbf{x}}(\mathscr{A})$, for typical Monte Carlo configurations, shows a phase transition. At higher temperatures, for {most} $\mathbf {x}$, the phase factors are close to one of the center elements $ e^{i 2\pi k/N}$, $k=0, \dots, N-1$. On the other hand,  below the transition, they are equally distributed on $Z(N)$, as a function of the spatial site $\mathbf{x}$. As a result, the Monte Carlo calculation  gives a transition from a non-vanishing to a vanishing  gauge-field average $  \langle P_{\mathbf{x}}  \rangle$, which is in fact $\mathbf{x}$-independent, where the {\it electric}  $Z(N)$ symmetry is not broken. This corresponds to a transition from a deconfined phase at higher $T$, where the quark free energy is finite, to a confined phase below $T_c$, where the free energy diverges. 

In the full Monte Carlo simulations, the relevance of  $Z(N)$ is also manifested in general  Wilson loops at asymptotic distances. In this regime, the string tension only  depends on the $N$-ality $k$ of $\rm{D}$, which determines how the center $Z(N)$ of $SU(N)$ is realized in the given quark representation~\cite{dFK}, 
\begin{gather}
{\rm D} \, (e^{i \frac{2\pi}{N}} I) = \left( e^{i \frac{2\pi }{N}} \right)^k I_{\mathscr D} \;.
\end{gather}

Regarding the confinement mechanism, lattice calculations aimed at determining the 
relevant degrees of freedom have been performed for many years. In particular,
procedures have been constructed to analyze Monte Carlo $U_\mu(x) \in SU(N)$  
link-configurations and extract center projected configurations $Z_\mu(x) \in Z(N)$~\cite{DelDebbio:1996lih, DelDebbio:1998luz,det1,det2} (for recent techniques to improve the detection of center vortices, see~\cite{fabrecent}). A given plaquette is then said to be pierced by a thin center vortex if the product of these center elements along the corresponding links is non-trivial. Observables may then be evaluated by considering vortex-removed and vortex-only configurations. The confining properties are only well described in the latter case~\cite{DelDebbio:1998luz,DelDebbio:1996lih,Engelhardt:1999xw,cp1,cp2,cp3,cp4,holl,lein2,lein3}. In the lattice, the analysis and visualization of center-vortex configurations~\cite{lein1} led to important insights regarding the origin of the topological charge density in the YM vacuum. In  3d (4d), thin center vortices are localized on worldlines (worldsheets) $\omega$. In this case, the Wilson loop in Equation \eqref{wloope} yields a center element 
\begin{gather}  
{\mathcal W}_{\rm D} ( {\cal C}_{\rm e})  = \mathcal{Z}_{\rm D}(\mathcal{C}_{\rm e}) = \frac{1}{\mathscr{D}} \, {\rm tr}\, \left[ {\mathrm D} \left(e^{i \frac{2\pi}{N}} I \right) \right]^{ {\rm L} (  \omega ,{\cal C}_{\rm e} )   }   \;, 
\label{lwl}
\end{gather}   
where ${\rm L}  (  \omega ,{\cal C}_{\rm e} )  $ is the total linking number between $\omega$ and ${\cal C}_{\rm e}$. This result also applies to thick center vortices, when their cores are completely linked by $\mathcal{C}_{\rm e}$. In this case,  {$\omega$} refers to the thick center vortex guiding centers. In the scaling limit, where the lattice calculations make contact with the continuum, the density of thin center vortices detected at low temperatures is finite 
\cite{Langfeld:1997jx, DelDebbio:1998luz}.  
Furthermore,  center vortices percolate and  have  positive stiffness~\cite{stiff1,stiff2}, while the fundamental Wilson loop average over $Z_\mu(x)$ displays an area law. This is in accordance with center-vortex condensation and the Wilson loop confinement criteria.   For $SU(2)$, a model based on the projected thin center-vortex ensemble captures  $97.7 \%$ of the fundamental string tension. On the other hand, the  percentage drops to $\sim$$62 \%$ for $SU(3)$~\cite{PhysRevD.69.014503}. One of the most important features of the center-vortex scenario is that  it naturally explains asymptotic $N$-ality:  the center element contribution in Equation \eqref{lwl} only depends on the $N$-ality of ${\rm D}$. For these reasons,  it is believed that the confinement mechanism should involve these degrees of freedom. For a recent discussion about this area of research, see \cite{greensite-book}. 

When it comes to accommodating the model-independent full Monte Carlo calculations, some questions arise.
 In 3d, the full asymptotic string tension dependence on D is very well fitted by the Casimir law~\cite{3dcasimirlaw}
 \begin{gather} \sigma^{(3)}_k  = \frac{k(N-k) }{N-1} \;, \label{3dcas}
 \end{gather} 
which is proportional to the lowest quadratic Casimir among those representations with the same $N$-ality $k$ of ${\rm D}$, which corresponds to the antisymmetric representation. 
 In addition, it is precisely at asymptotic interquark  distances where a model based on an ensemble of thin objects should be more reliable.  
 This is different at intermediate distances, where finite-size effects allowed for an explanation of the observed scaling with the Casimir of D~\cite{PhysRevD.57.2603,PhysRevD.75.034501}.
 Then, one question is: how to capture 
the asymptotic law  in  Equation \eqref{3dcas} from an average  over percolating thin center-vortices?    In 4d, where the available data cannot tell between a Casimir or a  Sine law~\cite{4dlaw}   
 \begin{gather} \sigma^{(4)}_k  = \frac{k(N-k) }{N-1} \makebox[.5in]{\rm vs.} \sigma^{(4)}_k = \frac{\sin {k\pi/N}}{\sin \pi/N}  \;, 
 \end{gather}  
is there any ensemble based on center-vortices that could reproduce one of these behaviors? More importantly, how can one explain this together with the formation of the confining flux tube observed in the  lattice? This means reproducing the L\"uscher term~\cite{L_scher_2002,at1,at2} and the observed transverse field distributions (see ~\cite{Cosmai-2017,Yanagihara2019210,Kitazawa,su3}, and 
references therein). Here, we shall review some developments  aimed at providing a possible answer to these questions. 

In Section \ref{cv-ens}, we shall discuss the simplest Abelian center-vortex ensembles. In Section \ref{abelian}, we
summarize, from different points of view, additional non-Abelian information and correlations that could be natural ingredients to be taken into account. In Section \ref{sec4}, we review ensembles of percolating oriented and non-oriented center vortices in 3d and 4d, their effective field description, as well as the possibility to accommodate the asymptotic properties of the confining string. Finally, in Section \ref{mixed}, we 
discuss recent lattice results in the light of our effective description,
and present some perspectives.

\section{ Center-Vortex Ensembles}
\label{cv-ens}

The idea that center vortices are the dominant degrees of freedom in the infrared regime means, in practice, that the Wilson loop average at asymptotic distances may well be captured by modeling the average of the center-elements in Equation \eqref{lwl}. 
This line of research was mainly explored in the lattice~\cite{randomsu2} by considering an ensemble of fluctuating worldlines (in 3d) or worldsurfaces (in 4d) with tension and stiffness  (see also the discussion at the beginning of Section \ref{4d-en}). For example, in 4d, a theory of fluctuating center-vortex worldsurfaces in four dimensions was introduced by considering the lattice action~\cite{randomsu2} 

\begin{align}
    S_{\rm latt}(\omega) = \mu \mathcal{A}(\omega) + c N_p\;,
\end{align}
where $\mathcal{A}(\omega)$ is the area of the vortex closed worldsurface $\omega$, formed by a set of plaquettes, and 
$N_p$ is the number of pairs of neighboring plaquettes of the surface lying on different planes. The latter term, as well as the lattice regularization, contribute  to the stiffness of the vortices. This model, initially introduced for $SU(2)$, and then generalized for $SU(3)$~\cite{randomsu3}, is able to describe important features, such as the confining string tension for fundamental quarks and the order of the deconfinement transition. This type of model can be also formulated in the continuum. The objective is the same, that is, looking for natural ensemble measures to compute center-element averages
and compare them with the asymptotic information extracted from 
the full Monte Carlo average $\langle \mathcal{W}_{\rm D}(\mathcal{C}_e) \rangle$. A successful comparison is expected to give important clues about the underlying mechanism of confinement. When computing center-element averages in the continuum, the simplest model has the form: 
\begin{align}
   \langle \mathcal{Z}_{\rm D}(\mathcal{C}_e) \rangle = \mathcal{N}\sum_{\omega} e^{-S(\omega)} \frac{1}{\mathscr{D}} \, {\rm tr}\, \left[ {\mathrm D} \left(e^{i \frac{2\pi}{N}} I \right) \right]^{ {\rm L} (  \omega ,{\cal C}_{\rm e} )   }\;, \label{generalidea} 
\end{align}
where $\sum_\omega$ represents the sum over different configurations in a diluted gas of closed worldlines (in 3d) or worldsurfaces (in 4d). The  
 weight factor $e^{-S(\omega)}$ implements 
 the effect of center-vortex tension ($\mu$) and stiffness ($1/\kappa$) observed in the lattice~\cite{stiff1,stiff2}. More precisely, $S(\omega)$ contains a term proportional to the length  or area  of $\omega$, and another one proportional to a power of the absolute value of the curvature of $\omega$. {See Equation \mbox{\eqref{stiff3d}} for an explicit formula in 3 dimensions.} $S(\omega)$ could also contain interactions with a scalar field $\psi$ that, when integrated with a corresponding weight $W(\psi)$, generates interactions among the variables $\omega$. 
 
 Extended models can also be introduced where the defining elements are not only given by $\omega$ but also by additional labels. At the level of the gauge field variables $A_\mu$, the center-vortex sectors can be characterized by different mappings $S_0\in SU(N)$ containing defects (see Section \ref{YM-ens}). A center vortex with guiding center $\omega$ and magnetic weight $\beta$ is 
  characterized by $S_0=e^{-i\chi \beta\cdot T}$, $\beta\cdot T\equiv \beta|_q T_q$, where $\chi$ is a multivalued angle that changes by $2\pi$ when going around $\omega$, and $T_q$, $q=1, \dots, N-1$ are the Cartan generators. 
  As they carry a single weight, these vortices are known as oriented (in the Cartan subalgebra). For elementary center vortices, the tuple $\beta $ is one of the magnetic weights $\beta_i$ ($i=1, \dots, N$) of the fundamental representation. In the region outside the vortex cores, $A_\mu$ is locally a pure gauge configuration constructed with $S_0$. Then, for fundamental quarks, the contribution to a large loop contained in that region is $i$-independent and given by the elementary center-element $(1/N)\, {\rm tr} \left( e^{- i2\pi \beta_i\cdot T} \right) = e^{ i2\pi/N}$ to the power ${\rm L} (  \omega ,{\cal C}_{\rm e})$. 
 Different elementary fluxes may join to form more complex configurations, provided this is done in a way that conserves the flux. For example,   {$N$ center-vortex guiding centers} associated with different magnetic weights $\beta_i$ can be matched{. For simplicity, let us consider the $SU(3)$ case in three dimensions and a configuration characterized by $S_0=e^{i\chi_1\beta_1\cdot T}e^{i\chi_2\beta_2\cdot T}$, where $\chi_1$ and $\chi_2$ are multivalued when going around the closed worldlines $\omega_1$ and $\omega_2$, respectively. These worldlines could meet at a point, then follow a common open line $\gamma$, and again bifurcate to close the corresponding loops. In this case, we would have a pair of fluxes entering the initial point, carrying the fundamental weights $\beta_1, \beta_2$, and a flux leaving along $\gamma$, carrying the weight $\beta_1+\beta_2$. In $SU(3)$, this sum is an antifundamental weight $-\beta_3$. In other words, there are three fluxes entering the initial point, which carry the three different fundamental weights $\beta_1, \beta_2, \beta_3$. This can be readily generalized to $SU(N)$, where $N$ fluxes carrying the different fundamental weights can meet at a point, as these weights satisfy $\sum_i \beta_i =0$.}  Vortices may also be non-oriented~\cite{Reinhardt:2001kf}, in the sense that they may not be described by a single weight. In this case, the center-vortex components with different fundamental weights are interpolated by instantons in 3d and monopole worldlines in 4d. These lower dimensional junctions, which carry a flux of the form $\beta_i -\beta_j$, should be weighted with additional phenomenological terms in $S(\omega)$. Furthermore, in the 4d case, three monopole worldlines carrying fluxes $\beta_i- \beta_j$, $\beta_j - \beta_k$, $\beta_k-\beta_i$ can be matched at a spacetime point. Similar higher-order matching rules are also possible. In what follows, we shall discuss the different ensembles, starting with the simplest possibilities in 3d and 4d.
 
\section{ Abelian Effective Description of Center Vortices}      \label{abelian}

In this section, we shall briefly discuss center-vortex ensembles formed by diluted closed worldlines in 3d (Section \ref{3den}) or worldsurfaces in 4d (Section \ref{4d-en}), characterized by no other properties than tension, stiffness, and vortex--vortex interactions. No additional degrees of freedom, matching rules or correlations with lower dimensional objects will be considered here. 
   
\subsection{Three Dimensions}   
\label{3den}
 
In a planar system, thin center vortices are localized on points, so they are created or annihilated by a field operator $\hat{V}(x)$. The emergence of this order parameter can be clearly seen by applying polymer  techniques  to center-vortex worldlines~\cite{deLemos:2011ww}. 
In ~\cite{Oxman-Reinhardt-2017},  the center-element average  for fundamental quarks, over all possible diluted loops, was initially represented in the form 
\begin{eqnarray}
\langle \mathcal{Z}_{\rm F}(\mathcal{C}_e) \rangle = \mathcal{N}   \int [D\psi]\, e^{-W[\psi]} \, e^{  \int_{0}^{\infty}\frac{dL}{L}\;  \int  dx \int du \,  Q(x,u,x,u,L)  }  \;,
\label{q-smooth} 
\end{eqnarray} 
 where $Q(x,u,x_0,u_0,L)$ is the
 integral over all paths with length $L$, starting (ending) at $x_0$ ($x$) with unit tangent vector $u_0$ ($u$), in the presence of  scalar and vector sources  {$\psi$} and $\frac{2\pi}{N} s_\mu$, and weighted by tension and stiffness. {The factor $W[\psi]=\frac{\zeta}{2}\int d^3x\, \psi^2(x)$ generates, upon integration of the auxiliary scalar field $\psi$, repulsive contact interactions between the loops with strength given by the parameter $\frac{1}{\zeta}$.}  Indeed, as in the exponential we have $x=x_0$, $u=u_0$, its expansion generates the diluted loop ensemble. As usual, the factor $1/L$
is to avoid loop overcounting when choosing $x_0$ on a given loop. The external source $s_\mu$ is localized on a surface  $S(\mathcal{C}_{\rm e})$ whose border is the Wilson loop. As a consequence, it generates the intersection numbers between the loop-variables in $Q$ and $S(\mathcal{C}_{\rm e})$, which coincide with the different linking-numbers. Using the large-distance behavior of $Q(x,u,x_0,u_0,L)$, which satisfies a Fokker--Planck diffusion equation (given by Equations \eqref{Qexp} and \eqref{app1}, with $b_\mu$ Abelian, and $D(\Gamma_\gamma[b_\mu])$ being {the} complex number $\Gamma_\gamma[b_\mu]$) we then showed that the ensemble average of center elements
becomes represented by a complex scalar field  $V(x)$, 
\begin{gather}
\langle \mathcal{Z}_{\rm F}(\mathcal{C}_e)  \rangle\approx
 \mathcal{N} \int [{\cal D}V][{\cal D}\bar{V}]\,   e^{-\int
	d^{3}x\,\left[ \frac{1}{3 \kappa}\,  \overline{D_\mu V} D_\mu V +\frac{1}{2\zeta}\, (\overline{V} V -v^2)^2  \right]} \;, \nonumber \\
	 v^2 \propto -\mu \kappa >  0 \makebox[.3in]{,}  D_\mu = \partial_{\mu}- i \frac{2\pi }{N} s_\mu 
	 \label{ese} \;.
\end{gather}   

\begin{figure}[H]     \centering
\begin{tabular}{cc}
\includegraphics[scale=.21]{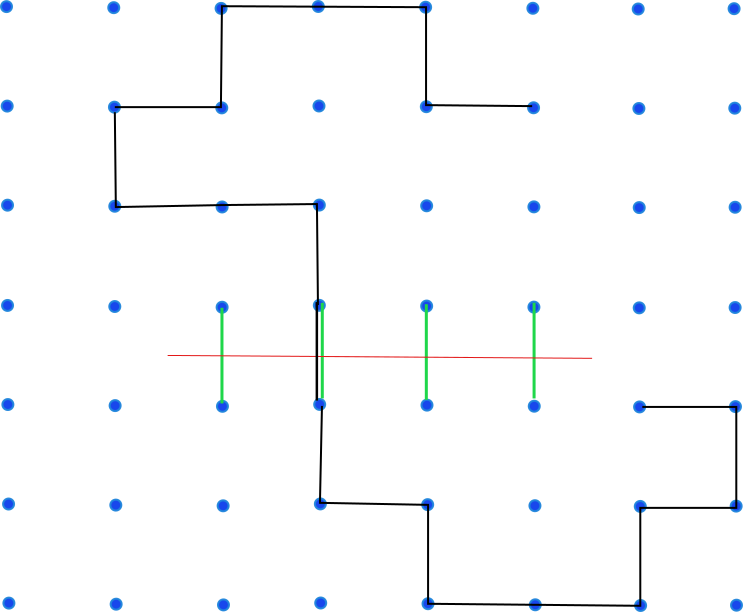} & \hspace{11mm}\includegraphics[scale=.21]{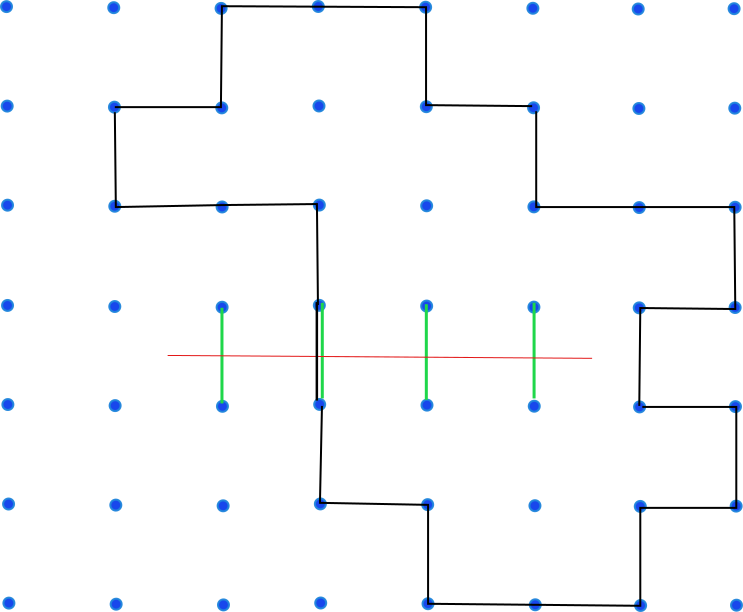} \\
(\textbf{a}) & \hspace{11mm}(\textbf{b}) \\
\end{tabular}  
\caption{The Wilson loop and the frustration are represented in red and green, respectively.  Configurations of type (\textbf{a}), which involve sites joined by open lines, do not contribute to the partition function. Only 
site configurations joined by loops, like the one in (\textbf{b}), contribute (with a center-element).  }    
\label{CV-link}     
\end{figure}

 This was obtained for small (positive) stiffness $1/\kappa$ and repulsive contact interactions. The scalar field $V$ is originated due to the approximate  behavior of $Q(x,u,x_0,u_0,L)$ in Equation \eqref{app-sol}, which turns the exponential in Equation \eqref{q-smooth} into a functional determinant. The squared mass parameter of this field is  proportional to
$ \kappa \mu$, where  $\mu$ is the center-vortex tension. For percolating objects ($\mu < 0$), the $U(1)$ symmetry of the effective field theory is spontaneously broken ($\kappa \mu < 0$). Among the consequences, we have:

\begin{enumerate}

\item  In the center-vortex condensate, 
 the effective description is dominated by the soft Goldstone modes, $V(x) \sim v\, e^{i\phi(x)}$. Then, the calculation of the center-element average is neither Gaussian nor dominated by a saddle-point, as it involves a compact scalar field $\phi$ and large fluctuations;

\item  This is better formulated in the lattice, where the Goldstone mode sector  is governed by a 3d XY model with frustration  
\begin{equation}
S^{(3)}_{\rm latt} = \tilde{\beta} \, \sum_{x, \mu }  \mathrm{Re} \left[ 1- 
 e^{i \gamma(x + \hat{\mu})} e^{-i \gamma(x)} e^{-i \alpha_\mu ({x})}   \right] \;.
 \label{xymodel}
\end{equation}   

The external source in Equation \eqref{ese} translates into the frustration $e^{ i \alpha_\mu (\mathbf{x})} = e^{i \frac{2\pi}{N}}$ if  $S
 ({\cal C}_{\rm e})$ is crossed by the link  and is trivial otherwise;

\item In the expansion of the partition function, due to the measure $ \prod_x \int_{-\pi}^{\pi} d\gamma(x) $,  the terms that contribute contain products of the composite $e^{ i \gamma(x + \hat{\mu})} e^{-i \gamma(x)}$ (or its conjugate) over links organized forming loops. Otherwise, the integrals over the site variables at the line edges vanish (see Figure \ref{CV-link});

\item Due to frustration, every time $\mathcal{C}_{\rm e}$ is linked, a center element is generated. Then, in the lattice, the closed center-vortex worldlines in the initial ensemble, which led to Equation \eqref{q-smooth} and gave origin to the effective description \eqref{ese}, are represented by the loops of item 3.

\end{enumerate} 

This point of view will be useful to propose other ensemble measures relying on lattice models, as in the case where the derivation of the effective description is not known, see for example Sections \ref{4d-en} and \ref{4den}. It is also interesting to see that the initial ensemble properties encoded in Equation \eqref{q-smooth} are recovered close to the 3d XY model critical point, as expected. Indeed, using the same techniques reviewed in~\cite{Klein-gf} for the case without frustration, the partition function may be formulated in terms of integer-valued divergenceless currents, originated after using the Fourier decomposition
\begin{align}
    e^{\beta\cos\gamma} =\sum_{b=-\infty}^\infty I_b(\beta)e^{ib\gamma}\;,
\end{align}
at every lattice link. The resulting expression turns out to be equivalent to a grand canonical ensemble of non-backtracking closed loops formed by currents of strength $|b_\mu|=1$. In the model without frustration, close to the critical point $\beta_{\rm c}  \approx 0.454 $ (continuum limit), the relevant configurations are known to be formed by large loops rather than by multiple small loops, and multiple occupation of links is disfavored, thus making contact with the initial properties parametrized in the ensemble (see  Table \ref{tab1} below).  

\begin{table}[H]    
\caption{ The correspondence between the effective field and the 3d XY model representations of the Abelian center$-$vortex ensemble.\label{tab1}}
\setlength{\tabcolsep}{10mm}
\begin{tabular}{|p{6.5cm}||p{5.5cm}|  }
\hline
\textbf{3d XY}    & \textbf{Effective Fields}  \\
\hline
 large loops are favored  & negative tension $\mu$    \\
  multiple small loops are disfavored &   positive stiffness $1/\kappa$   \\
 multiple occupation of links is disfavored & repulsive interactions \\
\hline
\end{tabular}
\end{table}

\subsection{Four Dimensions}   
\label{4d-en}
      
\begin{figure}[H]        \centering
\begin{tabular}{cc}
\includegraphics[scale=.18]{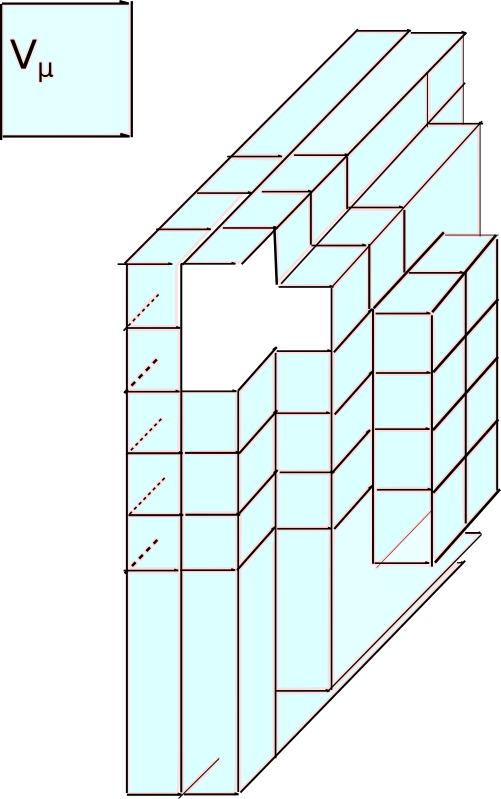} & \hspace{11mm}\includegraphics[scale=.18]{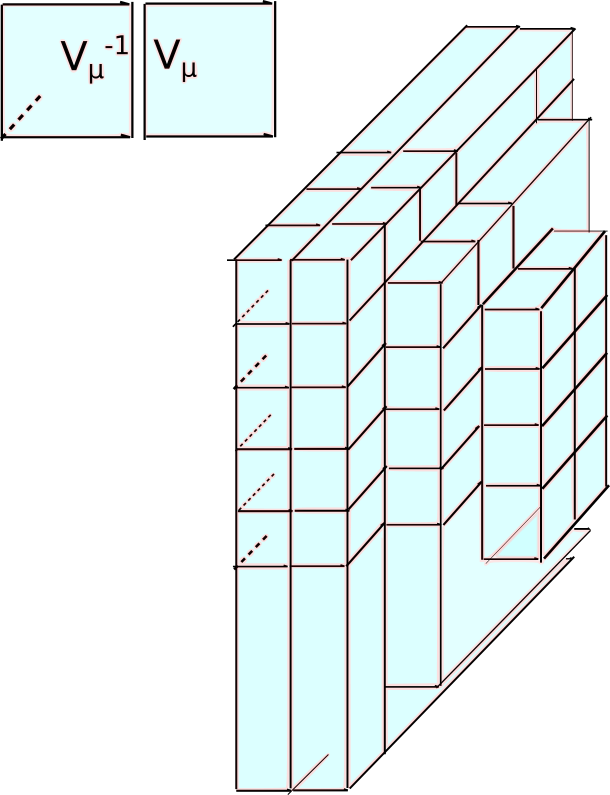} \\
(\textbf{a}) & \hspace{11mm}(\textbf{b}) \\
\end{tabular}  
 
\caption{(\textbf{a})
 Configurations formed by link variables distributed on plaquettes organized on an open surface 
 do not contribute, as the $V_\mu$ link$-$variables at the surface edges cannot form singlets; (\textbf{b})
 when they are organized on closed surfaces, singlets can be formed and the group$-$integral is non$-$trivial. }      
\label{lsur}        
\end{figure}

 Regarding the effective  description of 4d ensembles based on random surfaces, as in the $3~+~1$ dimensional world center vortices are one-dimensional objects spanning closed worldsurfaces, the emergent order parameter would be a {\it string field}. However, unlike the 3d case, a derivation starting from the ensemble of closed worldsurfaces with stiffness is still lacking. 
Such generalization should initially describe a growth process where a surface is generated, and then derive a Fokker--Planck equation for the lattice loop-to-loop probability. Similarly to what happens with end-to-end probabilities for polymers, where  stiffness is essential to get a 
continuum limit when the monomer size goes to zero~\cite{kleinert,fred}, curvature effects are expected to be  essential for the continuum limit of triangulated random surfaces. Indeed, ensembles of surfaces which consider only the Polyakov (or Nambu-Goto) action leads to a phase of branched polymers~\cite{ambjorn,wheater}. 
On the other hand, in ~\cite{Rey}, the phase fluctuations of an Abelian string field with frozen modulus were approximated by a lattice {\it field} theory: the $U(1)$ gauge-invariant Abelian Wilson action. In other words, the Goldstone modes for a condensate of one-dimensional objects are gauge fields.
 Motivated by this enormous simplification and by an analogy with the 3d case, in ~\cite{mixed} we proposed a Wilson action with frustration as a starting point to define a  measure  for percolating center vortices in four dimensions. This proposal will be discussed in Section \ref{4den}. For the time being, we summarize the main initial steps,  which are analogous to items 1--4 in Section \ref{3den}:

\begin{enumerate}

\item In the center-vortex condensate, 
 the effective theory is dominated by the soft Goldstone modes, which are represented by an emergent compact Abelian gauge field $V_\mu \in U(1)$. In the center-vortex context, we proposed another natural one based on $V_\mu \in SU(N)$ (see Section \ref{4den});   

\item The lattice version of the Goldstone mode sector is given by a Wilson action with frustration; 

\item  In the expansion of the partition function, the relevant configurations to compute the gauge model correspond to link-variables on the edges of plaquettes organized on closed surfaces (see Figure \ref{lsur});

\item The frustration is non-trivial on plaquettes $x,\mu,\nu$ that intersect  $S(\mathcal{C}_{\rm e})$.
Every time a closed surface links $\mathcal{C}_{\rm e}$, a center-element for quarks in the representation ${\rm D}$ is generated.

\end{enumerate}

Thus,  the main simplification in 4d is that,  in a condensate, the effective description can be captured by a local field. Similarly to 3d, where the soft modes can be read in the phase of the vortex field 
$V(x)\sim v\, e^{i\gamma(x)} $,   the natural soft modes in 4d are given by a compact gauge field, 
 \begin{gather}
V(C) \sim v\, e^{i\gamma_\Lambda(C)} \makebox[.3in]{,}
\gamma_\Lambda(C) = \oint_{C} dx_\mu \, \Lambda_\mu \;.
\end{gather}

\section{Center-Vortex Gauge Fields, Matching Rules, and Correlations }  \label{sec4}
  
The simplest  center-vortex ensembles discussed in Section \ref{abelian} could provide an important basis to understand the confinement mechanism at asymptotic distances. However, they do not contain enough ingredients to reproduce  more intricate properties. In this section, we shall discuss the center-vortex gauge fields and typically non-Abelian elements that could characterize the associated ensembles.

\subsection{Thick Center Vortices and Intermediate 
Casimir Scaling}

Before discussing generalized center-vortex ensembles with matching rules and non-oriented components, let us recall how the consideration of center-vortex thickness and the natural non-Abelian orientations in the gauge group can account for the observed  Casimir scaling at intermediate distances.  Some ideas along this line were initially pursued in ~\cite{cornwall}.
In ~\cite{PhysRevD.57.2603,PhysRevD.75.034501} (see also~\cite{Deldar_2001}), a simple model was put forward in the lattice, where the contribution to a planar Wilson loop along a curve $\mathcal{C}_{\rm e}$ was modeled. The starting point is to postulate an ensemble of thick center vortices whose total flux, as measured by a fundamental holonomy, have different possibilities $z^j = e^{i2\pi j/N}$,  $j=1, \dots, N-1$. When a thick center vortex is partially linked, the contribution to the Wilson loop is given by the insertion of a group element $G_j(x,S)$  that depends on the location ($x$) of the center-vortex midpoint (or guiding center) with respect to ${\mathcal C}_{\rm e}$. It also depends on a group orientation $S$, 
\begin{align}
    G_j(x,S)=S \mathcal{G}_j(x) S^\dagger  \;,
\end{align}
where $\mathcal{G}_j=\, \exp\left[i\,\alpha_j \cdot T\right]$ is in the Cartan subgroup and the tuples $\alpha_j$ are formed by model-dependent scalar profiles. These  profiles implement the natural condition that \mbox{$G_j(x,I) = z^j I_N$,} if the thick center vortex is fully enclosed by $\mathcal{C}_{\rm e}$, it is $I_N$ if it is not enclosed at all, and it gives an interpolating value otherwise.  After averaging over random group orientations in ~\cite{PhysRevD.57.2603,PhysRevD.75.034501}, they arrived at
\begin{align}
 &    \sigma_{\mathcal{C}_{\rm e}}({\rm D}) \equiv -\sum_x  \frac{1}{A}\ln (1-\sum_{j=0}^{N-1}f_j(1- \frac{1}{\mathscr{D}}{\rm Tr}\, {\rm D} \left(\,  \mathcal{G}_j\right)))\label{tension}\;,
\end{align}
where $f_j$ is the probability that a given plaquette of the planar surface enclosed by $\mathcal{C}_{\rm e}$ be pierced by the midpoint of a center-vortex of type $j$, $\sigma_{\mathcal{C}_e}(\rm{D})$ is the string tension in representation $\rm{D}$, and $A$ is the minimal area of $\mathcal{C}_{\rm e}$. At intermediate distances, after some natural approximations, an appropriate choice of profiles,
and using the key formula
\begin{align} 
    {\rm Tr}\,({\rm D}(T_q){\rm D}(T_p))= \mathscr{D} \,\delta_{qp}\frac{C_2({\rm D})}{N^2-1}\;,
\end{align} 
the Casimir Scaling
\begin{equation}
   \frac{\sigma_{\rm I}({\rm D})}{\sigma_{\rm I}({\rm F})} = \frac{ C_2({\rm D})}{C_2({\rm F})}\;
   \label{inter}
\end{equation} 
was obtained. In ~\cite{PhysRevD.57.2603,PhysRevD.75.034501},  based on a specific choice of probabilities and profiles, it was also possible to reproduce different asymptotic behaviors, such as the Casimir and the Sine law. In Section \ref{mixed},
we shall review a different line based on oriented and non-oriented center vortices, which naturally lead to an asymptotic Casimir law.
As these models are generated from weighted center-element averages, they are expected to be applicable in the asymptotic region.  

\subsection{Center-Vortex Sectors in Continuum YM Theory}
\label{YM-ens}

Center vortex correlations were considered for the first time in ~\cite{thooft}. In (2~+~1)d Minkowski spacetime, the order--disorder algebra in Equation \eqref{Vop3} says that the action of $\hat{V}({\bf x })$ on $|A\rangle $ gives
\begin{gather}
\hat{{\mathcal W}}_{\rm F} ( {\cal C}_{\rm e})  \left(  \hat{V}({\bf x }) |A\rangle \right) 
= e^{i \frac{2\pi}{N}}\,  {\mathcal W}_{\rm F} ( {\cal C}_{\rm e}) \left( \hat{V}({\bf x}) |A\rangle \right)  \;, 
\end{gather}  
if ${\bf x}$ is encircled by ${\cal C}_{\rm e} $, and it leaves the state $|A\rangle$ unaltered otherwise. Here, $|A\rangle $ is a state with well-defined shape in the Weyl gauge $A_0=0$, That is, $ \hat{V}({\bf x}) |A\rangle$ is a state where a thin center-vortex is created on top of $A_i$.  In particular, the action of $\hat{V}^N({\bf x})$ is trivial. 
Then, the possible phases were effectively described  by a model with magnetic $Z(N)$ symmetry
\begin{align}
    {\cal L} =\partial^\mu \bar{V}\, \partial_\mu V + m^2\, \bar{V} V + \frac{\lambda}{2} \, (\bar{V} V)^2 + \xi \, (V^N+\bar{V}^N)  \;. \label{thooft-mod} 
\end{align} 

 This  includes quadratic and quartic correlations, as well as the $N$-th order terms that capture the possibility that $N$ vortices may annihilate. 
The case $m^2 > 0$ would correspond to a Higgs phase where center vortices are in the spectrum of asymptotic states.  The case $m^2<0$ corresponds to a center-vortex condensate, with $N$ degenerate classical vacua, so that $Z(N)$ is spontaneously broken.  For a detailed analysis of this effective description, see ~\cite{kovner,koganreview}. In ~\cite{thooft}, based on the center-vortex operator definition $ \hat{V}({\bf x} )|A\rangle 
= |A^{S} \rangle $, discussed in Section \ref{intro}, 3d Euclidean vortex Green's functions $ \langle  \bar{V}(y) V(x) \rangle $ were defined. This was done by considering the YM path-integral over configurations $A_\mu$ with boundary conditions around  the pair of points $x,y \in \mathbb{R}^3$, such that a vortex is created at $x$, it is then propagated, and finally annihilated at $y$. When $|x-y| \to \infty$, an exponential decay would correspond to a Higgs phase and $\langle V \rangle =0$, because of the  clustering property. This agrees with the discussion above, where the Higgs phase $m^2>0$
is characterized by a $Z(N)$ symmetric vacuum.  On the other hand, a condensate would correspond to a Green's function that tends to a constant. 

Now,  from the definition of the operator $\hat{V}({\bf x})$, it is clear   that it introduces singularities in the gauge fields. If $A$ is smooth, the configuration $A^{S}$ is singular, with a field strength containing a delta-singularity at the center vortex location ${\bf x}$. 
As pointed out by 't Hooft, the operator's definition  could be made more precise by smearing the singularities 
over an infinitesimal region around ${\bf x}$. Otherwise, we would be working with singular infinite action gauge fields. Although this direction was not pursued in that work,  the smeared Green's functions could depend on the choice of boundary conditions, for the mapping $S \in SU(N)$, around $x$ and $y$. In other words, the vortex field $\hat{V}$ could hide non-Abelian degrees of freedom which are not evidenced by the algebra in Equation \eqref{Vop3}, which only depends on properties with respect to the Wilson loop. 

In ~\cite{Oxman:2015ira},
we proposed a partition of the full configuration space  of {\it smooth} gauge fields $\{ A_\mu \}$ into sectors $\mathcal{V}(S_0)\subset \{A_\mu\}$ characterized by topological labels $S_0$. For this objective, we introduced $N_{\rm f}$ auxiliary adjoint scalar fields $\psi_I$  by means of an identity in the YM path integral, which constrain them to be a solution
to a classical equation of motion for the minimization of an auxiliary action $S_{\rm aux}(\psi_I,A)$. Imposing regularity and boundary conditions, the solution $\psi_I(A)$ is unique, and can be decomposed by means of a generalized polar decomposition $\psi_I(A) = S q_I S^{-1}$, where $S(x)\in SU(N)$ and $(q_1, \dots, q_{N_{\rm f}})$ is a ``modulus'' tuple.  The phase defects cannot be eliminated by regular gauge transformations $U$, which act on the left 
$S \to US$. A gauge field is then said to belong to a given sector $\mathcal{V}(S_0)$ if $S(A)$ is equivalent to a  class representative $S_0$. The continuum of possible labels $S_0$ are characterized by the location of oriented and non-oriented 
center-vortex guiding centers, with all possible matching rules (see the discussion in Section \ref{cv-ens}). Although a possible label for an oriented center-vortex would be 
$S_0 =e^{i\chi\beta\cdot T}$, a typical non-oriented configuration is characterized by $S_0 =e^{i\chi\beta\cdot T} W$. In 3d, close to some points (instantons) on the center-vortex worldline generated by $e^{i\chi\beta\cdot T}$, the mapping $W$ behaves as a Weyl transformation 
that changes the fundamental weight $\beta$ to $\beta'$. Similarly, in 4d, the change occurs at some monopole worldlines on the center-vortex worldsurfaces generated by $e^{i\chi\beta\cdot T}$ (see ~\cite{mixed}).   
The full YM partition function and averages of observables were then represented by a sum over partial contributions, 
\begin{gather} 
   Z_{\rm YM}= \sum_{S_0} Z_{(S_0)} \makebox[.3in]{,}  \langle O \rangle_{\rm YM}=\frac{1}{Z_{\rm YM}} \sum_{S_0} \int_{\mathcal{V}(S_0)} [D\mathcal{A}_\mu] \, O\,  e^{-S_{\rm YM}}   \label{part0} \;.  
\end{gather}

Here, $\sum_{S_0}$ is a short-hand notation for the contribution originated from the continuum of labels $S_0$. These ideas provided a glimpse of a path connecting first principles Yang--Mills theory to an ensemble containing all possible center-vortex configurations. In addition to addressing this important conceptual issue, the partition into sectors may circumvent the well-known Gribov problem when fixing the gauge in non-Abelian gauge theories, as Singer's no go theorem~\cite{Singer:1978dk} only applies to global gauges in configuration space {(see \mbox{ \cite{copies}} for a detailed discussion)}. In ~\cite{Oxman:2015ira},  the gauge was locally fixed by a regular gauge transformation that rotates $S$ to the reference $S_0$, which is  imposed by a sector dependent condition $f_{S_0}(\psi)=0$. Furthermore, the theory was shown to be renormalizable in the vortex-free sector~\cite{PhysRevD.101.085007}. The extension of the renormalization proof to sectors labeled by center vortices is under way, and will be presented elsewhere.
An interesting consequence of this construction is that a new label may be generated by the right multiplication, $S_0\to S_0 \tilde{U}^{-1}$, with regular $\tilde{U}$, which is not necessarily connected to $S_0$ by a regular gauge transformation. That is, given a center-vortex sector, there is a continuum of physically inequivalent sectors characterized by  non-Abelian d.o.f. where the defects are located at the same spacetime points.  In the context of effective Yang--Mills--Higgs models, which describe the confining string as a smooth topological classical vortex solution, the presence of similar internal d.o.f. was previously noted in a large class of color-flavor symmetric theories~\cite{David,it,shif,David2,marshakov,PhysRevD.71.045010,new1,new2,new3,new4}.

\section{Mixed Ensembles of Oriented and Non-Oriented Center Vortices} 
\label{mixed}

The general properties of center vortices discussed so far motivate the search for a natural ensemble that captures all the  asymptotic properties of confinement. Among them, the formation of a confining flux tube is the most elusive one in this scenario. 
The formation of this object would also explain 
the L\" uscher term, which has not been observed in projected center-vortex ensembles.  Furthermore, 
the asymptotic Casimir law (cf. Equation \eqref{3dcas}) should be reproduced in 3d, while in 4d we would like to understand the coexistence of $N$-ality with the Abelian-like flux tube profiles~\cite{Cosmai-2017,Yanagihara2019210,Kitazawa,su3}. It is clear that a confining flux tube requires  an ensemble whose effective description
contains topological solitons, namely, a confining domain wall in (2~+~1)d and a vortex in 
(3~+~1)d. However, the simple models of oriented and uncorrelated center vortices discussed  in Section \ref{abelian} do not have the conditions to support these topological 
objects\footnote{Namely, a SSB pattern with discrete classical vacua in $(2+1)$d and multiple connected vacua in $(3+1)$d.}. In what follows, we shall review how the inclusion
of the center-vortex matching rules and correlations with lower dimensional defects (see \mbox{Sections \ref{cv-ens} and \ref{YM-ens}}) could fill the gap between center-vortex ensembles and the formation of a flux tube. 
In~\cite{refute,doublewilson}, lattice studies showed that the 4d Abelian-projected lattice is not represented by a monopole Coulomb gas, but rather by
collimated fluxes attached to the monopoles. In the continuum, these configurations correspond to the previously discussed non-oriented center vortices.  While in 4d the lower dimensional defects on center-vortex worldsurfaces are monopole worldlines, in 3d they are instantons. The relevance of non-oriented center vortices to generate a non-vanishing Pontryagin 
index was shown in ~\cite{Reinhardt:2001kf}.  Now, although oriented and non-oriented  
center vortices, located at the same place, would contribute to a large Wilson loop with the same center-element, it is natural to weight them with different effective actions. In the second case,  the measure should also depend on the location of the lower-dimensional defects. 

 \subsection{3d Ensemble with Asymptotic Casimir Law}

In this section, we review the mixed ensembles formed by oriented and non-oriented center-vortices with $N$-line matching rules introduced in ~\cite{O-S-D-3d}. In that reference,  to prepare the formalism so as to include the different correlations, we initially wrote the contribution to the Wilson loop of a thin center-vortex loop $l$ as
\begin{align}
  {\mathcal W}_{\rm D} ( {\cal C}_{\rm e})|_{\rm loop} =\frac{1}{N}   \, {\rm 
Tr}\, \Gamma_l[ b_\mu^{{\cal C}_{\rm e}}]  \makebox[.5in]{,} \Gamma_\gamma[b_\mu] = P\{e^{i\int_\gamma dx_\mu  b_\mu}\} \;,  \label{simplest}
\end{align}
where  $b_\mu^{{\cal C}_{\rm e}}=2\pi\beta_e\cdot T\, s_\mu^{{\cal C}_{\rm e}}$,  $\beta_e$ is the highest magnetic weight of ${\rm D}$, and $ s_\mu^{{\cal C}_{\rm e}}$ is a source localized on ${\cal C}_{\rm e}$.
Here, we use the notation $\beta_e\cdot T = \beta_e|_q T_q$, with $T_q$, $q=1\dots, N-1$ being the Cartan generators of $SU(N)$. Then, after weighting each loop with a phenomenological factor $e^{-S(l)} $  accounting for tension and stiffness (cf. Equation \eqref{generalidea}), and summing over
all possible diluted loops, we obtained the center-element average 
\begin{align} 
    \langle {\mathcal Z}_{\rm D} ( {\cal C}_{\rm e})  \rangle =  e^{\int_0^\infty\frac{dL}{L}\int dx\int du\,{\rm tr}\, 
Q(x,u,x,u,L)} \;, \label{zloops}
\end{align}
where $Q(x,u,x_0,u_0,L)$ is the integral over all the paths with length $L$ 
that begin at $x_0$ with unit tangent vector $u_0$, and end at $x$ with 
orientation $u$. This is given by Equation \eqref{Qexp}, using as ${\rm D}$ the fundamental representation.   
This object satisfies a non-Abelian diffusion equation whose large $\kappa$-limit (small stiffness)
solution (cf. Equation \eqref{app-sol}) led to
 approximate Equation \eqref{zloops}  by
\begin{align}
   \langle {\mathcal Z}_{\rm D} ( {\cal C}_{\rm e})  \rangle \approx Z_{\rm loops}= \mathcal{N}   \int[d\phi]\,e^{-\int d^3x\,\phi^\dagger 
\mathcal{O}\phi} \makebox[.5in]{,}
   \mathcal{O}= -\frac{1}{3\kappa}(I_N\partial_\mu-ib_\mu^{{\cal C}_{\rm e}})^2+\mu I_N \;,\label{loop}
\end{align}  
where  $\phi$ is an emergent complex scalar field in the fundamental representation.

One basic defining property of center vortices is that $N$ such 
objects can be virtually created out of the vacuum at $x_0$ and then annihilated at $x$. At the level of the gauge fields, this is related to the possibility of matching $N$ guiding centers each one carrying a different fundamental magnetic weight $\beta_i$, $i=1,\dots,N$, which  satisfy $\beta_1 + \dots + \beta_N  =0 $.
Then, to incorporate all possible oriented center-vortex line matchings (see Section \ref{Nmatch}), we  expanded the loop ensemble in Equation \eqref{loop} considering the $N$ types of weights, each one represented by a fundamental field $\phi_i$, $i=1, \dots , N$. At this point, the  center-element average over loops was generated from the
partition function 
\begin{align}
   & Z_{\rm loops}^N = \int[D\Phi^\dagger][D\Phi]\, e^{-\int d^3x \left[\frac{1}{3\kappa} {\rm 
Tr}\left((D_\mu\Phi)^\dagger D_\mu\Phi\right)+\mu {\rm Tr} 
(\Phi^\dagger\Phi)\right]}\;,
\end{align} 
where $\Phi$ is a complex $N \times N$ matrix with components $\Phi_{ij}=\phi_j|_i$.

\subsubsection{Including $N$-Vortex Matching}
\label{Nmatch}

The contribution to the Wilson loop of $N$
center-vortex worldlines  starting at $x_0$ and ending at $x$, and carrying different weights, was rewritten as
\begin{align}   
   {\mathcal W}_{\rm D} ( {\cal C}_{\rm e})|_{N-{\rm lines}} =\frac{1}{N!}\epsilon_{i_1\dots i_N}
    \epsilon_{i'_1\dots i'_N}\Gamma_{\gamma_1}[b_\mu^{{\cal C}_{\rm e}} ]|_{i_1i'_1}\dots 
\Gamma_{\gamma_N}[b_\mu^{{\cal C}_{\rm e}} ]|_{i_Ni'_N} \;. 
\label{Nline}
\end{align}

 By weighting each line in Equation \eqref{Nline} with the factor $e^{-S(\gamma_i)} $, and integrating over paths with fixed endpoints and over all the lengths $L_i$ (cf. Equation \eqref{Gf}), we obtained
\begin{align}
  C_N  \propto \int d^3x d^3 x_0  \, \epsilon_{i_1\dots i_N}\epsilon_{j_1\dots 
j_N}G(x,x_0)_{i_1 j_1}\dots G(x,x_0)_{i_N j_N}\;,
\label{CN}
\end{align} 
where $G(x,x_0)$ is the Green's function of the operator $\mathcal{O}$. In this manner, the $N$-line contribution in Equation \eqref{CN} and similar processes were generated by adding a term $\propto (\det\, 
\Phi+ \det\, \Phi^\dagger )  $. The effective description thus {obtained} is separately invariant under local and global  $SU(N)$ symmetries
$S_c(x), S_f \in SU(N)$
\begin{eqnarray}
  &&   \Phi\to S_c(x)\Phi \makebox[.4in]{,}b_\mu\to S_c(x)b_\mu 
S_c^{-1}(x)+iS_c(x)\partial_\mu S_c^{-1}(x) \;, \nonumber \\
  &&   \Phi\to\Phi S_f \;.
\end{eqnarray} 

In the effective description, other natural terms compatible with these symmetries, like the vortex--vortex interaction ${\rm Tr} (\Phi^\dagger\Phi)^2 $, should also be included,  thus leading to the center-element average $\langle {\mathcal Z}_{\rm D} ( {\cal C}_{\rm e}) \rangle = Z_{\rm v}[b_\mu^{{\cal C}_{\rm e}}]/Z_{\rm v}[0]$,
\begingroup\makeatletter\def\f@size{9}\check@mathfonts
\def\maketag@@@#1{\hbox{\m@th\fontsize{10}{10}\selectfont\normalfont#1}}
 \begin{align}
    Z_{\rm v}[b_\mu^{{\cal C}_{\rm e}}]=\int[D\Phi^\dagger][D\Phi]e^{-\int d^3x \left[\frac{1}{3\kappa} 
{\rm Tr}\left((D_\mu\Phi)^\dagger D_\mu\Phi\right)+\mu {\rm Tr} 
(\Phi^\dagger\Phi) + \frac{\lambda_0}{2} {\rm Tr} (\Phi^\dagger\Phi)^2 -\xi_0(\det\Phi+\det\Phi^\dagger)\right]} \;. 
\label{effd3}
\end{align}  \endgroup

This effective description has some similarities with 
the 't Hooft model (cf. Equation \eqref{thooft-mod}). More specifically, 
 they coincide for configurations of the type $\Phi=V 
I_N$.  However, there is no reason for the path-integral to favor this type of restricted configuration.  Up to this point, in the percolating phase ($\mu<0$), the quadratic and quartic terms tend to produce a manifold of  classical vacua labeled by $U(N)$, while 
the addition of the $\det \Phi$-interaction reduces this manifold to $SU(N)$. Then, 
unlike the 't Hooft model, in the SSB phase this effective description  has a continuum set of classical vacua  which precludes the formation of the stable domain wall. 
It is  interesting to formulate the Goldstone modes $V(x)\in SU(N)$ in the lattice, which leads to
\begin{equation} 
S^{(3)}_{\rm latt} (b_\mu^{\mathcal{C}_e})=  \tilde{\beta} \,\sum_{x, \mu} \mathrm{Re}\;  
 \left[ \mathbb{I} -   \bar{U}_\mu   V(x+\hat{\mu}) V^\dagger(x) )    \right]   \;, 
\end{equation} 
where $U_\mu(x) = e^{i2\pi \beta_{\rm e}\cdot T} \in Z(N)$, if the link $x,\mu$ crosses 
$S( {\cal C}_{\rm e})$, and it is the identity otherwise. As expected,   in the expansion of the partition function, besides the contribution of sites distributed on links that form loops, there is also one originated from $N$ lines that  start or end at a common site $x$. In the former case, the singlets are included in $N \otimes \bar{N}$, while in the latter they are in {the products of $N$ $V(x)$ or $V^\dagger(x)$} (compare with the Abelian case in Section \ref{3den}). In this way, the rules originating Equation \eqref{effd3} can be recovered in the lattice. This type of cross-checking is useful to better understand proposals of lattice ensemble measures in situations where it is harder to derive the effective field description, like in 4d~spacetime.

\subsubsection{Including Non-Oriented Center Vortices in 3d} 

In terms of Gilmore--Perelemov group  coherent-states  (see ~\cite{coherent1,coherent2} for a complete discussion or ~\cite{mixed} for a summary of the main ideas)  
$|g,\omega\rangle = g|\omega\rangle$, $g \in SU(N)$,
Equations \eqref{simplest} and \eqref{Nline} became
\begingroup\makeatletter\def\f@size{9}\check@mathfonts
\def\maketag@@@#1{\hbox{\m@th\fontsize{10}{10}\selectfont\normalfont#1}}
\begin{align}
   &   {\mathcal W}_{\rm D} ( {\cal C}_{\rm e})|_{\rm loop}  \propto \int d\mu(g)\, \langle 
g,\omega|\Gamma_l[b_\mu^{{\mathcal C}_{\rm e} } ]|g,\omega\rangle \label{repr} \;, \nonumber \\
   &     {\mathcal W}_{\rm D} ( {\cal C}_{\rm e})|_{N-{\rm lines}}   \propto \int d\mu(g)d\mu(g_0)\, \langle 
g,\omega_1|\Gamma_{\gamma_1}[b_\mu^{{\cal C}_{\rm e}}]|g_0,\omega_1\rangle\dots \langle 
g,\omega_N|\Gamma_{\gamma_N}[b_\mu^{{\cal C}_{\rm e}} ]|g_0,\omega_N\rangle\;.
\end{align}\endgroup

The first contribution can be thought of as associated to the creation of a center-vortex with initial fundamental weight $\omega$ and group orientation $g$, which is propagated along the closed worldline $l$, and is then annihilated. The second corresponds to $N$ vortices with different magnetic weights 
$\beta_i= 2N\, \omega_i$, $i=1,\dots,N$, created out of the vacuum at a spacetime point $x_0$, that follow 
separate worldlines $\gamma_i$ and then annihilate at $x_f$. Following a similar interpretation, and recalling that the center-vortex weights change at the instantons, we introduced   non-oriented center vortices. When a closed {object} is formed by n parts $ \gamma_1 \circ \gamma_2 \circ \dots \circ \gamma_n $ with $n$ instantons at points $x_1\dots x_n$, we considered the contribution
\begin{align}
    &C_n =\int d\mu(g_1)\dots\int d\mu(g_n) \langle g_1,\omega| 
g_2,\omega'\rangle \langle g_2,\omega| g_3,\omega'\rangle\dots \langle 
g_n,\omega| g_1,\omega'\rangle\times\\&\times \nonumber  \langle  
g_1,\omega'|\Gamma_{\gamma_n}[b_\mu^{{\cal C}_{\rm e}}]|g_n,\omega\rangle \dots \langle 
g_3,\omega'|\Gamma_{\gamma_2}[b_\mu^{{\cal C}_{\rm e}}]|g_2,\omega\rangle \langle 
g_2,\omega'|\Gamma_{\gamma_1}[b_\mu^{{\cal C}_{\rm e}}]|g_1,\omega\rangle\;. \label{instcorr}
\end{align}

Here, a center vortex is propagated along $\gamma_1$ from $x_1$, with orientation $g_1$ and weight $\omega$, up to $x_2$, with orientation $g_2$ and weight $\omega'$. At $x_2$, keeping the 
orientation $g_2$, the weight changes to $\omega'$, and then $\gamma_2$
is followed, etc. This precisely characterizes a non-oriented center vortex, where the flux orientation along the Cartan subalgebra changes.
Additionally, notice that   $|\omega'\rangle\langle \omega|$ is the root vector $E_\alpha$, which is in line with the presence of pointlike defects carrying adjoint charge. Moreover, when the chain configuration links the Wilson loop 
$\mathcal{C}_{\rm e}$, one of the holonomies $\Gamma_{\gamma_1},\dots,\Gamma_{\gamma_n}$ gives a center element, while all the others are 
trivial, thus leading to the expected center-element for a chain, up to a positive and real weight 
factor. Performing the integrals on the group, we arrived at 
an additional vertex and the final formula for the ensemble average of ${\mathcal W}_{\rm D} ( {\cal C}_{\rm e}) $, incorporating all the 
configurations discussed so far,  
\begin{align}
    &\langle {\mathcal Z}_{\rm D} ( {\cal C}_{\rm e}) \rangle = \frac{Z[b_\mu^{{\cal C}_{\rm e}}]}{Z[0]}\makebox[.5in]{,}Z[b_\mu]=\int 
[D\Phi]\,e^{-S_{\rm eff}(\Phi,b_\mu)}\;, 
\end{align}
\vspace{-12pt}
\begingroup\makeatletter\def\f@size{9}\check@mathfonts
\def\maketag@@@#1{\hbox{\m@th\fontsize{10}{10}\selectfont\normalfont#1}}
\begin{align}
    &S_{\rm eff}(\Phi,b_\mu) = \int d^3x\left({\rm Tr}(D_\mu\Phi)^\dagger 
D_\mu\Phi+V(\Phi,\Phi^\dagger)\right)\makebox[.5in]{,}D_\mu=\partial_\mu-i b_\mu 
\;,\\&
    V(\Phi,\Phi^\dagger)= \frac{3}{2}\lambda_0\kappa {\rm Tr} (\Phi^\dagger\Phi+\frac{\mu}{\lambda_0}
I_N)^2 -\xi_0 (3\kappa)^{\frac{N}{2}}(\det\Phi+\det\Phi^\dagger)-3\vartheta_0\kappa {\rm Tr}\left(\Phi^\dagger 
T_A\Phi T_A\right)\; \label{potential},
\end{align}\endgroup
where $\lambda_0,\xi_0,\vartheta_0>0$, and we have made the redefinition $\Phi \to \sqrt{3\kappa}\Phi $ of the field. When
vortices with positive stiffness percolate ($1/\kappa > 0$, $\mu < 0$), a condensate is formed. In the parameter region $\lambda_0\;,\xi_0\; >>\;\vartheta_0$, the most relevant fluctuations will be parametrized by $\Phi\propto S$, $S \in SU(N)$. It is interesting to check in the lattice how the different configuration types are recovered. The additional non-oriented component in the discretized theory is generated from the product of an adjoint variable arising from the new term
\begin{gather}
 {\rm Tr}\left(\Phi^\dagger 
T_A\Phi T_A\right)  \sim {\rm const.} \,  {\rm Tr}\left( {\rm Ad}(S)\right)  \;,  
\end{gather} 
 at a lattice site $x$,  with the adjoint contribution in $N\otimes\bar{N}$ associated with $V(x)$ and $V^\dagger(x)$.

\subsection{Saddle-Point Analysis in 3d} 

For non-trivial $\vartheta $, the $SU(N)$ classical vacua degeneracy is lifted, and the possible global minima become discrete:
 \begin{align}
    & \Phi=v\mathcal{Z}_N\makebox[.5in]{,}\mathcal{Z}_N=\{e^{i\frac{2\pi 
n}{N}}\;;n=0,1,\dots,N-1\}\;,\\&
    6\lambda_0\kappa N\left(v^2+\frac{\mu}{\lambda_0}\right)-2\xi_0(3\kappa)^{\frac{N}{2}} Nv^{N-2}-3\kappa\vartheta_0(N^2-1)=0\;. \label{vev}
 \end{align}
 
Thus, the presence of instantons opens the 
possibility of  stable domain walls that interpolate the 
different vacua. In this case, 
 the calculation may be approximated by   a saddle-point 
expansion.  Considering a large circular Wilson Loop $\mathcal{C}_{\rm e}$ centered at the origin of the $x_2-x_3$ plane, the effect of the 
source is simply to impose the boundary conditions
\begingroup\makeatletter\def\f@size{9}\check@mathfonts
\def\maketag@@@#1{\hbox{\m@th\fontsize{10}{10}\selectfont\normalfont#1}}
 \begin{align}
\lim_{x_1\to-\infty}\Phi(x_1,x_2,x_3)=vI_N\makebox[.3in]{,}\lim_{x_1\to\infty}
\Phi(x_1,x_2,x_3)=ve^{i2\pi\beta_e\cdot T}\makebox[.3in]{,} (0,x_2,x_3) \in S(\mathcal{C}_{\rm e})\;.
\label{boundarycond}
 \end{align}\endgroup
 
In ~\cite{O-S-D-3d}, we showed that the Ansatz
\begin{align} 
    \Phi=(\eta I_N+\eta_0\beta\cdot T)e^{i\theta\beta\cdot T}e^{i\alpha}\;
\end{align}
closes the equations of 
motion,  yielding scalar equations for the profiles 
$\eta,\eta_0,\theta,\alpha$. Due to the relation $e^{i2\pi\beta_e\cdot 
T}=e^{-i\frac{2k\pi}{N}}$, the boundary conditions \eqref{boundarycond} may be 
imposed either by a solution where $\alpha$ varies with $\theta$ constant, or 
vice versa. The first possibility is closely related to the 't Hooft model (cf. Equation \eqref{thooft-mod}). In the second case, the $\theta$ variation is governed by the Sine-Gordon equation
 \begin{align}
     \partial_{x_1}^2\theta=\frac{3\kappa\vartheta_0}{2}\sin(\theta)\;.
 \end{align}
 
In this manner, for quarks with $N$-ality $k$, we obtained the asymptotic Casimir Law  
 \begin{align}   
\epsilon_k = \frac{k(N-k)}{N-1}\epsilon_1\;,
 \end{align}
where $\epsilon_1$ is proportional to the Sine-Gordon parameter $3\kappa\vartheta_0$.  

\subsection{A 4d Ensemble with Asymptotic Casimir Law}
\label{4den}

Here, we review the ensembles of oriented and non-oriented center vortices in four dimensions as proposed in ~\cite{mixed}. In that study, instead of deriving the effective description of center-vortex ensembles 
with negative tension and positive stiffness, we started the discussion from the natural Goldstone modes defined on the lattice (see also Section \ref{4d-en}).  The missing steps are expected to be implemented by deriving diffusion loop equations including the effect of stiffness. The lattice description of an Abelian ensemble of worldsurfaces coupled to an external Kalb--Ramond field in the form
\begin{align}
&\int d\sigma_1d\sigma_2\, B_{\mu\nu}(X(\sigma_1,\sigma_2))\Sigma^{\mu\nu}(X(\sigma_1,\sigma_2))\makebox[.5in]{,}
\Sigma^{\mu\nu}=\frac{\partial X^\mu}{\partial\sigma_1}\frac{\partial X^\nu}{\partial\sigma_2}-\frac{\partial X^\nu}{\partial\sigma_1}\frac{\partial X^\mu}{\partial\sigma_2}\;,  
\label{cKR}
\end{align}
where $X^\mu(\sigma_1,\sigma_2)$ is a parametrization of the worldsurface, 
was obtained in ~\cite{Rey}. This was done in terms of a complex-valued string field $V(C)$, where $C$ is a closed loop formed by a set of lattice links. The associated action is 
\begin{align} 
&S_V= -\sum_C\sum_{p\in \eta(C)} \left[  \bar{V}(C+p)U_pV(C)+\bar{V}(C-p)\bar{U}_pV(C)\right]    + \sum_C m^2\bar{V}(C)V(C)\;. \label{lattstring}
\end{align}

 $\eta(C)$ is the set of plaquettes that share at least one common link with $C$,
while  $C+ p$ is the path that follows $C$ until the initial site of the common link, then detours through the other three links of $p$, and continues along the remaining part of $C$. In addition, the \mbox{coupling \eqref{cKR}} originates the plaquette field $U_p=e^{i a^2 B_{\mu\nu}(p)}$.  Then,  the following 
polar decomposition was considered
\begin{align}
    V(C)= w(C) \prod_{l\in C} V_l   \makebox[.5in]{,}   V_l  \in U(1) \label{decomp}\;,
\end{align}
with a phase factor that has a ``local'' character, as it was written in terms of the holonomy along $C$ of gauge field link-variables $V_l$. Finally, when a condensate is formed ($m^2< 0$), it was argued that the modulus is practically frozen\footnote{Similarly to the 3d case, this phase should be stabilized by a quartic interaction.}, so that $w(C)\approx w>0$. By using this fact in Equation \eqref{lattstring}, the only links whose contribution do not cancel are those  belonging to 
$p$:
\begin{align} 
    &\bar{V}(C+p)U_pV(C)=  w^2  \prod_{l\in C+p} \prod_{l'\in C} \bar{V}_{l}   U_p V_{l'}  
    =w^2 U_p \prod_{l \in p} \bar{V}_{l} \;. 
\end{align}

Thus,
\begin{equation} 
S^{(4)}_{\rm latt} (\alpha_p)=  \tilde{\beta} \,\sum_{p} \mathrm{Re}\;  
 \left[ \mathbb{I} -   \bar{U}_p   \prod_{l \in p} V_{l}    \right]   \;. \label{Abe4} 
\end{equation} 
where the sum is over all plaquettes $p$ and a constant was added such that the action vanishes for a trivial plaquette. Then, the description of a loop condensate, where loops are expected to percolate, is much simpler than that associated with a general phase. The string field parameter gives place to simpler gauge field Goldstone variables $V_\mu = e^{i\Lambda_\mu(l)}$, governed by a  Wilson action with frustration $U_p$. 
This was the starting input used in ~\cite{mixed}. An external Kalb--Ramond field 
that generates the center elements when the simplest center-vortex worldsurface link  $\mathcal{C}_{\rm e}$ 
is obtained by replacing  $B_{\mu \nu} \to  \frac{2\pi k}{N} \, s_{\mu \nu} $, 
where $k$ is the $N$-ality of the quark representation ${\rm D}$ and  
 \begin{align}
    &s_{\mu\nu}=\int_{S(C_e)}d^2\tilde{\sigma}_{\mu\nu}\delta^{(4)}(x-X(\sigma_1,\sigma_2)) \;,\\&
    d^2\tilde{\sigma}_{\mu\nu}=\frac{1}{2}\epsilon_{\mu\nu\alpha\beta}\left(\frac{\partial X^\alpha}{\partial\sigma_1}\frac{\partial X^\beta}{\partial\sigma_2}-\frac{\partial X^\beta}{\partial\sigma_1}\frac{\partial X^\alpha}{\partial\sigma_2}\right)d\sigma_1d\sigma_2
\end{align} 
is localized on $S(\mathcal{C}_{\rm e})$. In the lattice, this localized source  corresponds to a frustration $U_p = e^{i \alpha_p}$, where $\alpha_p = -2\pi k/N$ if $p$ intersects $S(\mathcal{C}_{\rm e})$ and it is trivial otherwise. Similarly to the 3d case, we can check a posteriori that the lattice expansion involves an average of center elements over closed worldsurfaces (see Section \ref{4d-en}).   
This is a consequence of the properties of $U(1)$ group integrals. This also applies to the non-Abelian extension $V_\mu \in SU(N)$, governed by  
\begin{equation}  
S^{\rm latt}_{\rm V} (\alpha_{\mu \nu}) =  \tilde{\beta} \,\sum_{\mathbf{x}, \mu < \nu } \mathrm{Re}\;  {\rm tr} \left[ I - \bar{U}_{\mu \nu} V_\mu(x) V_\nu(x + \hat{\mu}) V^\dagger_\mu(x + \hat{\nu}) V^\dagger_\nu(x)   \right] \;,  \nonumber 
\end{equation}  
where plaquettes are denoted   as usual.  
The closed surfaces are generated because $N \otimes \bar{N}$ contain a singlet. Interestingly, the $SU(N)$ version has additional configurations where $N$ open worldsurfaces meet at a loop formed by a set of links. 
This is due to the presence of a singlet in the product of $N$ link variables.  
Therefore, the associated normalized \mbox{partition function} 
\begin{gather}
\frac{Z_{\rm v}^{\rm latt} [\alpha_{\mu \nu}]}{Z_{\rm v}^{\rm latt} [0]}  \makebox[.5in]{,} Z_{\rm v}^{\rm latt} [\alpha_{\mu \nu}] = 
 \int [{\cal D} V_\mu] \, e^{-S^{\rm latt}_{\rm V}(\alpha_{\mu \nu}) } 
\end{gather} 
is an average of the center elements generated  when a Wilson loop in representation  ${\rm D} $ is linked by an ensemble of oriented center-vortex worldsurfaces with matching rules.

\subsection{Including Non-Oriented Center Vortices in 4d}  

\begin{figure}[H] \centering
\begin{tabular}{ccc}
\includegraphics[scale=.17]{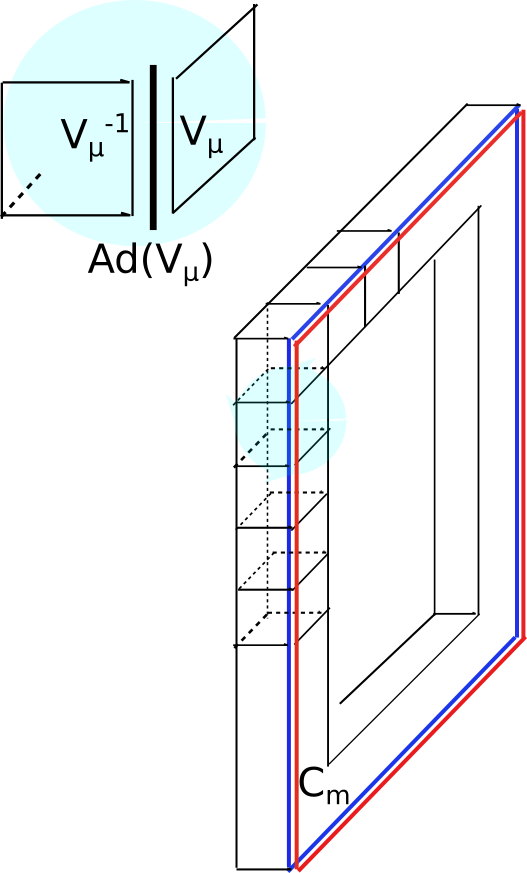} & \hspace{11mm}\includegraphics[scale=.17]{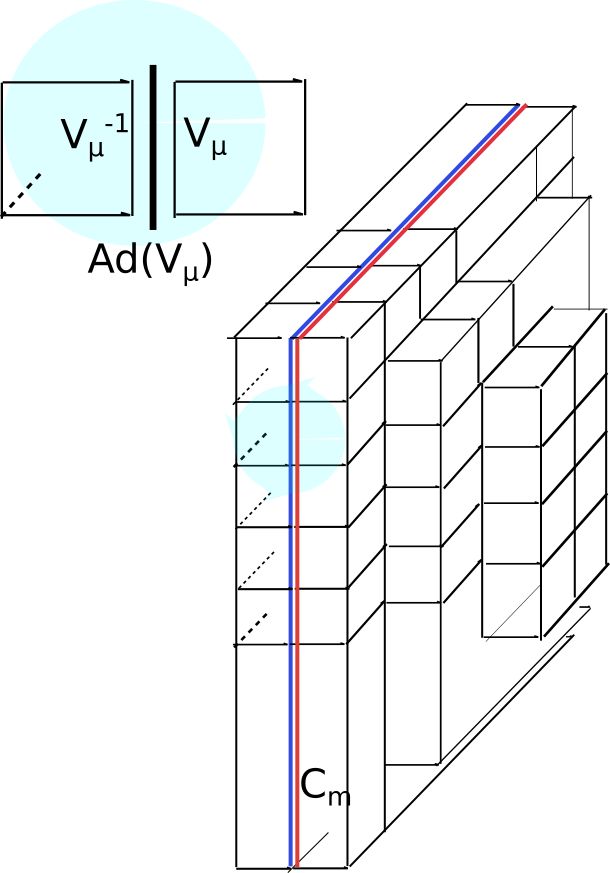} & \hspace{11mm} \includegraphics[scale=.17]{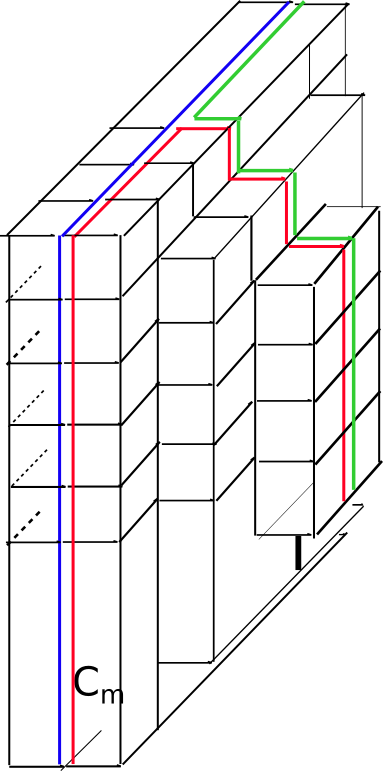}\\
(\textbf{a}) & \hspace{11mm}(\textbf{b}) & \hspace{11mm}(\textbf{c})\\
\end{tabular} 
\caption{Non$-$oriented center vortices containing monopole worldlines. We show a configuration that contributes to the lowest order in $\tilde{\beta}$ (\textbf{a}), and one that becomes more important as $\tilde{\beta}$ is increased (\textbf{b}). A non-oriented center vortex with three matched monopole worldlines is shown in (\textbf{c}).   }      
\label{mon-curr}    
\end{figure}

 Although thin oriented or non-oriented center vortices contribute with the same center-element to the Wilson loop, they are distinct gauge field configurations, with different Yang--Mills action densities. It is then important to underline that the ensemble measure could depend on the monopole component. In order to attach center vortices to monopoles, 
we  included  dual adjoint holonomies defined on a ``gas'' of monopole loops and fused worldlines. In this case, because of the integration properties in the group there are additional relevant configurations like those of Figure \ref{mon-curr}a,b.  
The use of adjoint holonomies 
 is in line with the fact that monopoles carry weights of the adjoint representation (the difference of fundamental weights), see ~\cite{mixed,Oxman:2015ira}. 

Then, partial contributions with $n$-loops were generated by
\begin{gather}
Z^{\rm latt}_{\rm mix} [\alpha_{\mu \nu}]\big|_{\rm p} \propto  \int [{\cal D} V_\mu] \, e^{-S^{\rm latt}_{\rm V}(\alpha_{\mu \nu})  } \,  {\mathcal W}^{(1)}_{\rm Ad} \dots  {\mathcal W}^{(n)}_{\rm Ad} 
\nonumber \\  
{\mathcal W}^{(k)}_{\rm Ad}    =  \frac{1}{N^2-1} \, {\rm tr}\,   \Big( \prod_{(x,\mu)\in \,
{\cal C}^{\rm latt}_k } {\rm Ad} \big( V_{\mu}(x) \big)   \Big)   
\label{w-latt} \;. 
\end{gather} 

\begin{figure}[H]    \centering
{\includegraphics[scale=.3]{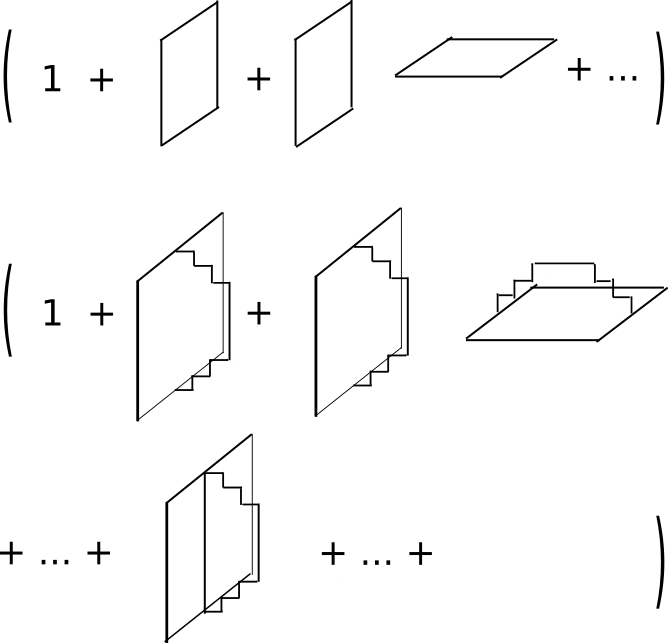} } 
\caption{\small Natural combinations of holonomies that can be used to model the mixed ensemble of oriented and non-oriented center vortices. Each contribution is weighted with tension and stiffness.  }    
\label{holoc}      
\end{figure}

In addition to the matching rules of $N$ worldsurfaces, which in the continuum occur as
$N$ different fundamental magnetic weights add to zero, monopole worldlines carrying different adjoint weights (roots) can also be fused. For example, when $N \geq 3$, three worldlines carrying different roots that add up to zero can be created at a point.   For this reason, we also considered  partial contributions to the ensemble like 
 \begin{gather}
 Z^{\rm latt}_{\rm mix} [\alpha_{\mu \nu}]\big|_{\rm p} \propto  \int [{\cal D} V_\mu] \, e^{- S^{\rm latt}_{\rm V}(\alpha_{\mu \nu})   } \, D^{\rm latt}_3  \;,
 \label{tres}
\end{gather} 
where $ D^{\rm latt}_3$ is formed by combining three adjoint holonomies ${\rm Ad}(\Gamma^{\rm latt}_j) $ (see Figure \ref{mon-curr}c). Other natural rules involve the matching of four worldlines. Then, weighting the monopole holonomies with the simplest geometrical properties (tension and stiffness),  the lattice mixed ensemble of oriented and non-oriented center vortices with matching rules can be pictorially
represented as 
 \begin{gather}
 Z^{\rm latt}_{\rm mix} [\alpha_{\mu \nu}] =   \int [{\cal D} V_\mu] \, e^{- 
 S^{\rm latt}_{\rm V} (\alpha_{\mu \nu})   } \, \times \dots
\label{mixZ}
\end{gather}
where the dots represent possible combinations of holonomies as illustrated in Figure \ref{holoc}.  

Then, noting that $ e^{i 2\pi k/N} = e^{-i\, 2\pi\, \beta \cdot w_{\rm e}}$, where $\beta$ is a fundamental magnetic weight and $w_{\rm e}$ is a weight of the quark representation ${\rm D}$, we considered the naive continuum limit, $ V_\mu(x) = e^{ia \Lambda_\mu(x)}$, $ \Lambda_\mu \in \mathfrak{su}(N) $,  
\begin{equation}
  Z_{\rm mix} [s_{\mu \nu}] = \int [{\cal D}\Lambda_\mu] \,   e^{-\int d^4x\,    \frac{1}{4\tilde{g}^2} \,   \left( F_{\mu \nu}(\Lambda ) - 2\pi s_{\mu \nu} \beta_{\rm e} \cdot T  \right)^2  } \, \times \dots
\end{equation}   

The dots represent all possible monopole configurations to be attached to center-vortex worldsurfaces (see Figure \ref{contC}). Each contribution was obtained using the methods in the Appendix.  The first factor in Figure \ref{contC} (monopole loops) generates emergent adjoint fields coupled to the effective field $\Lambda_\mu$. 
\vspace{-10pt}

\begin{figure}[H]     \centering  
{\includegraphics[scale=.32]{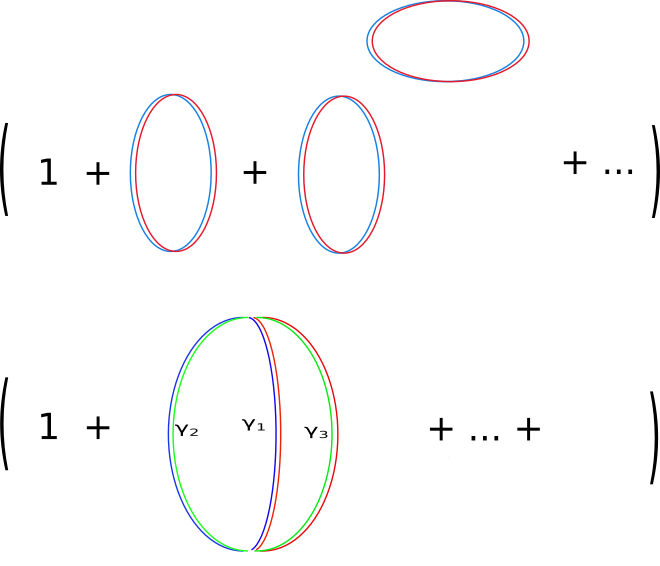}  }  
\caption{\small Continuum limit of the monopole sector. The worldline contributions are obtained from the solution to a Fokker--Planck diffusion equation.}  
\label{contC}
\end{figure}  
 For example, a diluted ensemble of a given species of monopoles, with tension $\tilde{\mu}$ and stiffness $\frac{1}{\tilde{\kappa}}$, is generated by

\begin{eqnarray}
  e^{  \int_{0}^{\infty}\frac{dL}{L}\;  \int  d^4x \, du \,  {\rm tr}\, Q(x,u, x,u,L)  } \;,
  \label{cont1}
\end{eqnarray}   
where $Q$ is given by Equation \eqref{Qexp} and ${\rm D}$ corresponds to the adjoint representation. 
In the small-stiffness approximation, the non-Abelian diffusion equation for $Q$  is solved by Equation \eqref{app-sol}, with 
\begin{eqnarray}  
 O = - \frac{\pi}{12 \tilde{\kappa}} \, \left( \partial_{\mu} - i\, {\rm Ad} \big(\Lambda_{\mu} \big) \right)^2  
 + \tilde{\mu}I_{\mathscr{D}_{\rm Ad}} \;. \label{Oad}
\end{eqnarray}  

Therefore, the factor in Equation \eqref{cont1} was approximated by
\begin{eqnarray}
&  & e^{\, - {\rm Tr} \,  \ln O  }
 = \int  [\mathcal{D}\zeta][\mathcal{D}\zeta^\dagger]   \,     e^{- \int d^4x\,  
\left(    ( D_\mu \zeta^{\, \dagger} , 
 D_\mu \zeta ) + \tilde{m}^2   (\zeta^\dagger, \zeta )  \right)}
\nonumber \\
&&  \tilde{m}^2 = (12/\pi) \,  \tilde{\mu} \tilde{\kappa}  \makebox[.5in]{,}  D_\mu(\Lambda)\, \zeta  = \partial_{\mu} \zeta - i\, [\Lambda_{\mu} ,  \zeta ] \;,
\end{eqnarray} 
where $\zeta$ is an emergent complex adjoint field, and we have introduced the {Killing} 
product between two Lie algebra elements $X,Y$ as $( X,Y) \equiv {\rm tr} ({\rm Ad}(X) {\rm Ad}(Y))$. 
 In the continuum, the path-integral of $ {\rm Ad}(\Gamma[\Lambda])$ over shapes and lengths led to the Green's function for the operator $O$, so that fusion rules like the one in Equation \eqref{tres} became effective Feynman diagrams. Indeed, 
 to differentiate the monopole lines that can be fused, the monopole loop ensemble was extended to include different species. At the end, 
 a set of real adjoint fields $\psi_I \in \mathfrak{su}(N)$ emerged
 ($I$ is a flavor index). This, together with the non-Abelian Goldstone modes (gauge fields), led to a class of effective Yang--Mills--Higgs (YMH) models, 
 \begin{eqnarray}
  Z_{\rm mix} [s_{\mu \nu}] =  \int [{\cal D}\Lambda_\mu] [{\cal D }\psi ] \,   e^{-\int d^4x\,  \left[  \frac{1}{4g^2} \,   \left( F_{\mu \nu}(\Lambda ) - 2\pi s_{\mu \nu} \beta_{\rm e} \cdot T  \right)^2 +  \frac{1}{2} (D_\mu \psi_I , 
 D_\mu \psi_I ) +  V_{\rm H}(\psi)  \right]  } \;.
\end{eqnarray}

The vertex couplings weight the abundance of each fusion type.  Percolating monopole worldlines (positive stiffness and negative tension) favor a spontaneous symmetry breaking phase that can easily correspond to S$U(N) \to Z(N)$ SSB. This pattern has been extensively studied in the literature (see ~\cite{David,it,shif,David2,marshakov,PhysRevD.71.045010,Fidel2,Auzzi-Kumar,conf-qg} and references therein).

\subsection{Analysis of the Saddle Point in 4d}  

In ~\cite{O-V,O-S,O-S-J}, we investigated a possible model  containing $N^2-1$ real adjoint scalar fields $\psi_I$ and ${\rm Ad}(SU(N))$ flavor symmetry,
\begin{equation}
\label{potentialhiggs}
V_{\rm H}(\psi) = c+\frac{\mu^2}{2}( \psi_A,\psi_A) + \frac{\kappa}{3}f_{ABC}( \psi_A \wedge \psi_B, \psi_C) + \frac{\lambda}{4}( \psi_A\wedge \psi_B)^2\;,
\end{equation}
where $X\wedge Y \equiv -i[X,Y]$. This model includes some of the correlations previously discussed. The case $\tilde{\mu}=0$ is specially interesting. At this point, the 
 classical  vacua are
\begin{subequations}
\begin{align}
 \Lambda_\mu = \frac{i}{g}S\partial_\mu S^{-1}    \makebox[.5in]{,}   \psi_A = v S T_A S^{-1}   \;.
\end{align}
\end{subequations}

 Then, the Higgs vacua manifold is ${\rm Ad}(SU(N))$ and the system undergoes $SU(N)\rightarrow Z(N)$ SSB, which leads to stable confining center strings. Interestingly, at $\tilde{\mu}=0$, we were able to find a set of BPS equations that provide vortex solutions whose energy is
\begin{gather}
\epsilon = 2\pi \tilde{g} v^2 \beta\cdot2\delta  \;,
\label{ener-law}
\end{gather}
where $\delta$ is the sum of all positive roots of the Lie algebra of $SU(N)$.
Using an inductive proof based on the Young tableau properties, we showed that the smallest  $\beta\cdot2\delta $ factor is given by the $k\text{-}{\rm A}$ weight, the highest weight of the totally antisymmetric representation with $N$-ality $k$. Then, for a general representation ${\rm D}(\cdot)$ with $N$-ality $k$, the asymptotic string tension satisfies
\begin{equation}
   \frac{\sigma({\rm D})}{\sigma({\rm F})} = \frac{ C_2({k\text{-}{\rm A}})}{C_2({\rm F})} = \frac{k(N-k)}{N-1}\;,
   \label{casim-l}
\end{equation} 
which is one of the possible behaviors observed in lattice simulations. 
Furthermore, the radial energy distribution transverse to the string  
is $k(N-k)$ times the distribution for a Nielsen--Olesen vortex. For $k=1$, this agrees with the YM energy distribution of the fundamental confining string, recently obtained from lattice Monte Carlo simulations~\cite{Cosmai-2017,Yanagihara2019210,Kitazawa,su3}.

\section{Discussion}   

We reviewed ensembles formed by oriented and non-oriented center vortices in 3d and 4d Euclidean spacetime that could capture the confinement properties of $SU(N)$ pure Yang--Mills theory.   Different measures to compute center-element averages were discussed. In 3d and 4d, they include percolating oriented center-vortex worldlines and worldsurfaces that generate emergent Goldstone modes, which correspond to compact scalar and gauge fields, respectively. The models also have the natural matching rules of $N$ center vortices, as well as the non-oriented component where  center-vortex worldlines (worldsurfaces) are attached to lower-dimensional defects, i.e., instantons (monopole worldlines) in 3d (4d). 
In addition to the  weighting center vortices with tension and stiffness,  it is also natural to include additional weights for the lower dimensional defects. In 4d,  monopole matching rules are also included.   The corresponding effective field content and the  SSB pattern may lead to the formation of a confining center string, represented by a domain wall (vortex) in two-dimensional (three-dimensional) real space.  The L\" uscher term is originated as usual, from the string-like transverse fluctuations of the flux tube. An asymptotic Casimir law can also be accommodated. This 
asymptotic behavior was observed in $3$d, while in $4$d it is among the possibilities.

More recently, the transverse distribution of the 4d YM energy-momentum tensor  $T_{\mu \nu}$ and the field profiles have been analyzed at intermediate and nearly asymptotic distances~\cite{Cosmai-2017,Kitazawa,Kondo,su3}.
In ~\cite{Kitazawa}, it was numerically shown that the $T_{\mu \nu}$ tensor of the Abelian Nielsen--Olesen (ANO) model cannot fit the $SU(3)$ data at the vortex guiding center for $L=0.46$ fm (intermediate distance) and $L=0.92$ fm (near asymptotic distance) at the same time. In fact, in ~\cite{su3}, it was shown that the  components of the energy-momentum tensor at the origin may not be accommodated for $L=0.46$ fm. Then, on this basis, 
an ANO effective model to describe the fundamental string was discarded. However, while it is clear that an effective model for the confining flux tube should work at asymptotic distances, it is not that obvious that the same model could be extrapolated to intermediate distances. By intermediate distances we mean those where the string tension scales with the quadratic Casimir of the quark representation. In particular, this is the region where adjoint quarks are still confined by a linear potential, before the breaking of the adjoint string. On the other hand, in the asymptotic region,  gluonic excitations around external quarks in a given irreducible representation
 $D(\cdot)$ may be created, so as to produce an asymptotic scaling law that only depends on the $N$-ality of $D(\cdot)$.
As discussed in this review, the  effective field descriptions were derived by considering the (weighted) average of center elements over oriented and non-oriented center vortices, which is expected to be applicable at asymptotic distances. 
 In other words, we wonder if it is meaningful to discard possible effective models on the basis of the lack of adjustment to lattice data on a wide range that includes the intermediate region, where these models are not expected to fully capture the physics. Additionally, note that the known mechanism to explain intermediate Casimir scaling is based on including center-vortex thickness. In turn, these finite-size effects are not included in the ensemble definition that leads to our effective model. Interestingly, while the lattice data  rule out the ANO model at intermediate distances $L=0.46$ fm, such profiles are still among the possibilities at the nearly asymptotic distance $L=0.92$ fm. Accordingly, the 4d $SU(N) \to Z(N)$ models we discussed in this review have
a point in parameter space where the infinite flux tube profiles Abelianize, while keeping all the required $N$-ality properties.  Additionally, the ideas presented in this review imply that not only an asymptotic Casimir law should be observed, but also that  the transverse confining flux tube profiles for quarks in different representations should be the same,  up to the asymptotic scaling law. This is true for both $3$d and $4$d, with the profiles being of the Sine-Gordon type in $3$d. It would be interesting to test these predictions with lattice simulations.

\vspace{6pt} 

\section*{Acknowledgements}
The Conselho Nacional de Desenvolvimento Cient\'{\i}fico e Tecnol\'{o}gico 
(CNPq), the Coordena\c c\~ao de Aperfei\c coamento de Pessoal de N\'{\i}vel 
Superior (CAPES), and the Deutscher Akademischer Austauschdienst (DAAD) are acknowledged for their financial support.

\appendix
\section{Non-Abelian Diffusion}
\label{diffu}

Center vortices in 3 dimensions and monopoles in 4 dimensions are propagated along worldlines in Euclidean spacetime. Then, the corresponding ensembles will naturally involve the building block $Q$ associated to a worldline with length $L$ that starts at
$x_0$ with orientation $u_0$ and ends at $x$ with final orientation $u$. This is given 
by
\vspace{-12pt}
\begin{align}
&Q(x,u,x_0,u_0,L)=\int[dx(s)]_{x_0,u_0}^{x,u} \, e^{-S(\gamma)}\,  {\rm D} \big( \Gamma_\gamma[b_\mu] \big)   \label{Qexp} \;, 
    \\&   \Gamma_\gamma[b_\mu] = P\{e^{i\int_\gamma  dx_\mu b_\mu}\}  \label{holono} \;,
\end{align}
where $S(\gamma)$ is a vortex effective action, and  an interaction with a 
general non-Abelian gauge field $b_\mu$ was considered. 
We are interested in the specific form 
\begin{gather}
S(\gamma) = \int_0^L ds\,  \left( \frac{1}{2\kappa} \dot{u}_\mu\dot{u}_\mu+\mu \right)  \makebox[.4in]{,}  u_\mu(s) = \frac{dx_\mu}{ds}   \;, \label{stiff3d}
\end{gather}
 which corresponds to tension $\mu$ and stiffness $1/\kappa$. These objects were extensively studied in ~\cite{mixed, GBO}. In what follows, we review the results obtained.
 
For the simplest center-vortex worldlines in 3d,  ${\rm D}$ is the defining $SU(N)$ representation,
while for monopole worldlines in 4d, ${\rm D}$ corresponds to the adjoint. 
   To derive a diffusion equation 
for this object, the paths were discretized into $M$ segments of length 
$\Delta L=L/M$. In this case, the path ordering was obtained from
\begin{align}
    P\{e^{-\int_0^L ds H(x(s),u(s))}\}=e^{-H(x_M,u_M)\Delta L}\dots 
e^{-H(x_1,u_1)\Delta L},
\end{align}
where $H(x,u)=-i {\rm D} (u_\mu  b_\mu(x))$. The 
relation between {the building block $Q_M$ associated to a discretized path containing $M$ segments of length $\Delta L$} and that associated with a path of 
length $L-\Delta L$ is given by:

\begin{align}
    &Q_M(x,u,x_0,u_0,L)=\int d^n x'd^{n-1}u'e^{-\mu\Delta L}\psi(u-u')\times 
\nonumber\\&e^{-\mu\Delta L}e^{-H(x,u)\Delta L}\delta(x-x'-u\Delta 
L)\, Q_{M-1}(x',x_0,u',u_0)\;, \label{chapman}
\end{align}
with 
\begin{align}
    \psi(u-u')=\mathcal{N}e^{-\frac{1}{2\kappa}\Delta L\left(\frac{u-u'}{\Delta 
L}\right)^2}
\end{align}
arising from the discretization of the stiffness term. It acts like an angular 
distribution in velocity space, which tends to bring $u'$ close to $u$. 
Expanding Equation \eqref{chapman} to first order in $\Delta L$, and taking the limit 
$\Delta L\to 0$, the diffusion equation
\begin{align}\label{app1}
    \left(\partial_L-\frac{\kappa\sigma}{2}\hat{L}_u^2+\mu+u_\mu(\partial_\mu-i 
{\rm D} (b_\mu)\right)Q(x,u,x_0,u_0,L)=0\;, 
\end{align}
was obtained,
to be solved with the initial condition
\begin{align}&
    Q(x,u,x_0,u_0,0)=\delta(x-x_0)\delta(u-u_0)I_{\mathscr{D}}\;. \label{diffeqs}
\end{align}

 $\mathscr{D}$ is the dimension of the quark representation D and $\hat{L}_u^2$ is the Laplacian on the sphere $S^{n-1}$. The constant $\sigma$ is given, in $n$ spacetime dimensions, by
\begin{align}    
    &\sigma=\frac{\sqrt{\pi}}{2^{n-3}}\frac{\Gamma\left(\frac{n-2}{2}
\right)\Gamma\left(\frac{n+1}{2}\right)}{\Gamma^2\left(\frac{n-1}{2}
\right)\Gamma\left(\frac{n-3}{2}\right)}\left(\frac{4\Gamma(n-3)}{
\Gamma\left(\frac{n-3}{2}\right)}-\frac{\Gamma(n-1)}{\Gamma\left(\frac{n+1}{2}
\right)}\right)\;.
\end{align}

 For the cases considered in this review ($n=3,4$), $\sigma=1,2/\pi$, 
respectively. In the limit of small stiffness, there is 
practically no correlation between $u$ and $u_0$, which allowed for a consistent 
solution of these equations with only the lowest angular momenta components:
\begin{align}
    &Q(x,u,x_0,u_0,L)\approx Q_0(x,x_0,L) \makebox[.5in]{,}\partial_L 
Q_0(x,x_0,L)=-O Q_0(x,x_0,L)\;, \\&
    O=-\frac{2}{(n-1)\sigma\kappa 
n}(\partial_\mu-iD(b_\mu))^2+\mu\makebox[.5in]{,} 
Q_0(x,x_0,0)=\frac{1}{\Omega_{n-1}}\delta(x-x_0)\;,
\end{align}
$\Omega_{n-1}$ being the solid angle of $S^{n-1}$. This implies,
\begin{align}
    &Q(x,u,x_0,u_0,L)\approx \langle x|e^{-L \mathcal{O}}|x_0\rangle\;.
    \label{app-sol}
\end{align}

Then, in this limit, we also have 
\begin{gather}
 \int_0^\infty dL\, du \, du_0  \int [Dx]_{x_0,u_0}^{x,u} \, e^{-S(\gamma)}\,    {\rm D}(\Gamma[b])=  \int_0^{\infty} dL\, du \, du_0\,  Q(x,u,x_0,u_0,L)  \nonumber \\
\approx  \langle x| O ^{-1} |  x_0 \rangle   \makebox[.3in]{,} O\, G(x,x_0) = \delta(x-x_0  ) \, I_{\mathscr{D}} \;.
\label{Gf}
\end{gather}

  \end{document}